%% file: Main.tex
\documentclass[lettersize,journal]{IEEEtran}
\usepackage{amsmath,amsfonts}
\usepackage{algorithm}
\usepackage{algcompatible}
\usepackage{array}
\usepackage[caption=false,font=normalsize,labelfont=sf,textfont=sf] {subfig}
\usepackage{textcomp}
\usepackage{stfloats}
\usepackage{url}
\usepackage{verbatim}
\usepackage{graphicx}
\usepackage{cite}

\usepackage{caption}
\usepackage{color,soul}
\usepackage{float}
\usepackage{tikz}
\usepackage{pgfplots}
\usepackage{pgfplotstable}
\pgfplotsset{compat=1.18}
\usepackage{pgf-pie}
\usetikzlibrary{patterns}
\usepackage[dvipsnames]{xcolor}
\usetikzlibrary{plotmarks}
\usepackage{multirow}
\usepackage{makecell} 
\usepackage{booktabs}

\usepackage{etoolbox}
\usepackage[colorlinks=true, linkcolor=blue, citecolor=blue, urlcolor=blue]{hyperref}

\makeatletter
\renewcommand{\@cite}[2]{%
  \leavevmode
  \begingroup
  \def\@temp##1{%
    \fcolorbox{green}{white}{##1}%
  }%
  [\@for\@cite@temp:=#1\do{%
    \@temp{\@cite@temp}%
  }]%
  \endgroup
}
\makeatother
\makeatletter
\let\old@ref\ref
\newcommand{\figboxref}[1]{\fcolorbox{cyan}{white}{Figure~\ref{#1}}}
\newcommand{\tabboxref}[1]{\fcolorbox{orange}{white}{Table~\ref{#1}}}
\newcommand{\algboxref}[1]{\fcolorbox{magenta}{white}{Algorithm~\ref{#1}}}

\makeatother

\begin{document}

\title{LLM-as-a-Judge for Contest-Style Human-AI Co-Creation: A Reliability-Aware Evaluation Framework for Coding} 


\author{Md Faizul Ibne Amin,~\IEEEmembership{Member,~IEEE,} Yutaka Watanobe,~\IEEEmembership{Member,~IEEE,} Daniel M. Muepu,~\IEEEmembership{Graduate Student Member,~IEEE,} Haruto Suzuki, Kenta Nanaumi, and Md Mostafizer Rahman

\thanks{Md Faizul Ibne Amin, Yutaka Watanobe, Daniel M. Muepu, Haruto Suzuki, and Kenta Nanaumi are with the Department of Computer and Information Systems, The University of Aizu, Aizu-Wakamatsu, Fukushima 965-8580, Japan (e-mail: aminfaizul007@gmail.com, yutaka@u-aizu.ac.jp, mdaniel108@gmail.com, lingmuwenda93@gmail.com, and m5291053@u-aizu.ac.jp).}

\thanks{Md Mostafizer Rahman is with the Lucy Family Institute for Data \& Society, University of Notre Dame, IN 46556, USA (e-mail: mostafiz26@gmail.com, mrahman3@nd.edu).}
}




\maketitle

\begin{abstract}
Large language models now judge open-ended outputs and co-create code with programmers, yet judgments of \emph{human-AI co-creation} must be reliable, comparable across models, and interpretable over multi-turn interaction. This work presents a rubric-driven \emph{LLM-as-a-Judge} framework for contest-style \emph{human-AI co-creation} in coding, evaluated on 15 participants, 13 problems, 184 trajectories, and 517 attempts from an AI-permitted track run in parallel with a programming contest. Four judges are evaluated under grouped cross-validation with cluster-bootstrap 95\% intervals and probed by injecting single defects into accepted programs against behavior-preserving controls. Three findings concern the judges as instruments and likely extend beyond this setting. First, the judge input is a first-order design variable: supplying the problem statement in place of an identifier rebuilds the judges' ranking (rank correlation 0.2617 [0.123, 0.391] for the three judges whose model was unchanged) and raises their ROC-AUC by 0.130 [0.061, 0.202] to 0.295 [0.214, 0.381]. Second, schema compliance does not imply scale validity: the judges emit as few as 8 distinct probability values, leaving as few as 6 of 99 thresholds effective. Third, accuracy and defect sensitivity are separable: Claude and Gemini are indistinguishable on accuracy (ROC-AUC 0.6985 [0.633, 0.763] against 0.7691 [0.719, 0.817]) yet separate on defect detection (0.8380 [0.806, 0.871] against 0.7512 [0.719, 0.783]). For co-creation, Success@Turn rises from 0.6033 [0.533, 0.674] at the first attempt to 0.8370 [0.783, 0.886] by the tenth, 44 of the 73 trajectories unsolved at first (60.3\%) succeed later, and revision size is unrelated to judged improvement (Spearman 0.0007 [$-$0.085, +0.090]).
\end{abstract}

\begin{IEEEkeywords}
LLM as a Judge, Human-AI co-creation, Large Language Models, AI for code, Software Engineering.
\end{IEEEkeywords}

\section{Introduction}
\label{Introduction}
\IEEEPARstart{R}{ecent} Transformer-based large language models (LLMs)~\cite{vaswani2017attention, rahman2025large}, pretrained at scale and instruction-tuned, follow natural-language (NL) specifications and produce controllable outputs~\cite{ouyang2022training, brown2020language}. In software engineering (SE) and programming, code-capable LLMs generate, summarize, explain, and refine solutions and support iterative debugging from NL prompts~\cite{crupi2025effectiveness, chen2021evaluating}, increasingly as partners in \emph{human-AI co-creation}, where artifacts are built jointly through multi-turn interaction, explicit constraints, and mixed initiative~\cite{rezwana2023designing, nguyen2024human, hassany2024human, kadenhe2025human}. Evaluation then becomes harder: co-creative traces contain partial attempts, revisions, and explanations, and the target extends beyond final correctness to constraint satisfaction, clarity, and usefulness~\cite{honarvar2025question, liu2023g, chang2024survey}. These challenges motivate scalable evaluation protocols such as \emph{LLM-as-a-Judge}, in which an LLM assesses open-ended artifacts under explicit criteria, reducing reliance on slow and inconsistent human grading~\cite{gu2024survey, li2024llms}.

In coding and SE, evaluation has traditionally relied on outcome-only signals such as compile success and pass/fail tests, which miss the value of intermediate artifacts such as partially correct patches, debugging hypotheses, and explanations~\cite{liu2023your, li2023generative}. Co-creation therefore needs reasoning-aware evaluation that captures product and process quality consistently and auditably~\cite{wang2025can, ji2025causality, koutcheme2025evaluating}. This study examines the problem in an AI-permitted co-creation track run in parallel with the official PC Koshien (PCK) Finals, where participants attempt the same released problems under the same time constraints while allowed to use AI tools~\cite{PCK}. Co-creative traces are operationalized as attempt-level instances grouped by $(\text{participant}, \text{problem})$ to avoid trajectory leakage, and judges receive the problem statement and the submitted code under three input conditions that vary only in the participant context supplied, returning schema-constrained rubric outputs.

LLM judging is commonly implemented as pairwise preference for ranking systems or as direct assessment under task-specific rubrics~\cite{zheng2023judging, liu2023g, pathak2025rubric, honarvar2025question}, and specialized open judge models target reproducibility~\cite{wang2024pandalm, zhu2023judgelm}. Judge outputs can vary with prompt wording, formatting, and presentation order, and can favor longer responses, particular positions, or outputs close to the judge's own distribution~\cite{chiang2023closer, dubois2024length, wang2025mcts, shi2025judging, wang2025assessing, wataoka2024self, ye2025learning}; judging also can be manipulated, and bias profiles differ between human and LLM evaluators~\cite{chen2024humans}. These concerns weigh more in coding and SE, where tasks involve long contexts, complex code changes, and realistic benchmarks such as SWE-bench~\cite{jimenez2023swe, jiang2025codejudgebench, tan2024judgebench}. Judging co-creative programming artifacts therefore requires protocols that are auditable and reliability-aware, meaning consistent under repeated use and demonstrably valid for what they measure~\cite{chang2024survey, li2023generative}.

Although \emph{LLM-as-a-Judge} has become a practical solution for scalable evaluation, existing research has largely emphasized judge bias, benchmark alignment, or general judge design rather than standardizing the end-to-end measurement pipeline needed for \emph{human-AI co-creation} artifacts in coding and SE specifically~\cite{shi2025judging, wataoka2024self, chen2024humans, wang2025assessing, tan2024judgebench, jiang2025codejudgebench}. Co-creative artifacts are heterogeneous and iterative, interleaving NL with code, partial revisions, and evolving constraints, so conclusions depend on judge reliability and on execution-level failures such as malformed outputs~\cite{chiang2023closer, dubois2024length}. A practical gap therefore remains, not for another pipeline but for evidence that one's judgments are valid measurements: a framework that yields auditable decisions instead of free-form critiques, is robust to invalid outputs, and supports reliability-aware reporting for trajectory-level co-creation analysis.

To fill this gap, we propose a unified rubric-driven \emph{LLM-as-a-Judge} framework for contest-style \emph{human-AI co-creation} in coding and SE, in which multiple judges (OpenAI, DeepSeek, Gemini, and Claude) return schema-constrained criterion-level outputs that are validated and repaired when necessary, and are analyzed through complementary metrics for discrimination, thresholded decision quality, probabilistic reliability, and inter-judge agreement. While existing pipelines produce judgments, this study also measures whether those judgments function as valid measurements, in a setting where an objective external verdict and the full interaction process are both available, and reports where they hold and where they fall short. A controlled perturbation study, which injects single defects into contest-accepted programs, supplies ground truth we control. In parallel, the same framework supports trajectory-based analysis of \emph{human-AI co-creation} through turn-wise rank consensus of the judges, Success@Turn, time-to-success, revision churn, and code-aware similarity. The empirical part is a first demonstration in this contest-style setting: claims about participants rest on that setting, whereas findings about the judges as instruments (the effect of input specification, the gap between schema compliance and scale validity, and the separation of accuracy from defect sensitivity) concern LLM judging more generally.

\noindent \textbf{Key points of this research.}
\begin{itemize}
  \item \textbf{Framework:} We introduce a rubric-driven \emph{LLM-as-a-Judge} measurement pipeline for contest-style \emph{human-AI co-creation} in coding and SE, emphasizing auditable outputs, schema validation and repair, and reliability-aware evaluation across multiple judge models.
 
  \item \textbf{Co-creation Dynamics:} We model human-AI interaction through trajectory-level signals including turn-wise progress, Success@Turn, and time-to-success, together with revision churn and code-aware similarity.
 
  \item \textbf{Evaluation:} A multi-metric evaluation suite is provided to quantify (i) discrimination (ROC-AUC and PR-AUC), (ii) thresholded decision quality (MCC), (iii) probabilistic reliability (LogLoss, Brier score, and ECE), (iv) inter-judge agreement (Cohen's $\kappa$ and Fleiss' $\kappa$), (v) revision behavior (NED and CodeBLEU), and (vi) perturbation-based validity (arm discrimination, detection specificity, and severity gradient).
 
  \item \textbf{Impact:} The resulting methodology provides a reproducible evaluation template that links process-aware co-creation analysis with reliability-aware judge evaluation for future AI-assisted coding and SE studies.
\end{itemize}

\section{Related Work}
\label{RW}
\paragraph{LLMs, AI-Assisted Programming, and Human-AI Co-Creation}
LLMs are studied as programming assistants in education and SE, for code generation, explanation, classification, refinement, and debugging under NL interaction~\cite{li2023generative, amin2024multi, amin2025source, chen2021evaluating, crupi2025effectiveness, rahman2026codet5, wu2025humanevalcomm, amin2026error}. Attention has moved from final outputs toward workflows in which users iteratively inspect, revise, and build on model-produced artifacts~\cite{liu2023g, honarvar2025question, hwang202580}, and studies of \emph{human-AI co-creation} describe these workflows as collaborative processes shaped by mixed initiative, iterative feedback, and human oversight, not one-shot automation~\cite{rezwana2023designing, nguyen2024human, hassany2024human}.

\paragraph{LLM-as-a-Judge for Evaluation}
Recent surveys~\cite{gu2024survey, li2024llms} describe how LLM judges are designed and evaluated across tasks, within the broader landscape of LLM evaluation~\cite{chang2024survey}. Within this view, Croxford et al.~\cite{croxford2025automating} benchmarked a medical \emph{LLM-as-a-Judge} framework against the validated PDSQI-9 instrument for clinical summarization, with GPT-o3-mini reaching an ICC of 0.818 at 22 seconds per evaluation on average. Kim et al.~\cite{kim2024prometheus} introduced an open evaluator supporting direct scoring and pairwise ranking under user-defined criteria. Wang et al.~\cite{wang2025can} compared \emph{LLM-as-a-Judge} methods with human scores on SE tasks and found output-based large LLM judges best aligned for code translation and generation. Crupi et al.~\cite{crupi2025effectiveness} found GPT-4-turbo the strongest of several judges for code generation and summarization, though it still made non-trivial misjudgments. SE-Jury~\cite{zhou2025se} treats five evaluation strategies as independent judges and dynamically selects a team among them to score the correctness of generated software artifacts across code generation, program repair, and code summarization. Code-DiTing~\cite{yang2025code} distills the reasoning of a large model into compact evaluators (1.5B and 7B parameters) that judge whether generated code is functionally consistent with its problem description, without references or test cases. Another line of work has focused on judge bias and reliability. Wataoka et al.~\cite{wataoka2024self} showed that LLM judges can favor their own or more familiar outputs, while Zhou et al.~\cite{zhou2024fairer} highlighted sensitivity to preference biases such as position and verbosity.

\paragraph{Human-AI Co-Creation in Coding and SE}
Hassany et al.~\cite{hassany2024human} proposed an authoring system in which an LLM generates line-by-line explanations of instructor-provided problems and code for humans to review and edit. In a controlled experiment with 109 participants, Wang et al.~\cite{wang2024rocks} found that ChatGPT sped up coding puzzles but helped less on broader development tasks. Ma et al.~\cite{ma2024teach} introduced HYPOCOMPASS, a tutoring system in which novice students guide an LLM-simulated learner through debugging. Daryanto et al.~\cite{daryanto2026human} showed in a within-subjects study that shared and personal AI support in triadic programming improves collaborative learning and social presence over human-AI baselines, and Basque et al.~\cite{EURECOM+8548} found human-LLM teaming in software reverse engineering helpful in some respects and harmful in others, through a practitioner survey and a controlled study.

\paragraph{Evaluation of Programming Performance, Reliability, and Multi-Rater Measurement}
A review of 121 studies on automated grading and feedback tools~\cite{messer2024automated} finds most systems oriented toward functional correctness and unit tests, with little attention to maintainability, readability, or documentation. Inter-coder agreement coefficients, particularly Krippendorff's $\alpha$, have been formalized to strengthen transparency in collaborative SE analysis~\cite{gonzalez2023reliability}. Wang et al.~\cite{wang2025axiom} introduced a multilingual benchmark for meta-evaluating code-focused LLM judges and reported strong ranking performance together with weaknesses in consistency and refinement-effort judgment. Student programs have been modeled as progression trajectories over intermediate code states, whose classes predict future behavior better than simpler baselines~\cite{morshed2021progression}, and Babe et al.~\cite{babe2024studenteval} built a benchmark of 1,749 student-written prompts to examine how code LLMs handle non-expert NL-to-code descriptions, including prompt ambiguity and inter-rater reliability.

\paragraph{Research Gap and Positioning of This Study}
These lines of work have advanced largely in parallel: co-creation studies foreground interaction behavior or task effectiveness, while judge-oriented work focuses on artifact-level evaluation, benchmark alignment, or general judge design, so the intersection of trajectory-level co-creation and reliability-aware \emph{LLM-as-a-Judge} evaluation in coding and SE has received little integrated treatment. This study addresses that intersection, examining process-aware co-creation behavior and multi-metric judge reliability and validity in a single contest-style setting, as an empirical bridge between the two lines. Relative to recent surveys~\cite{gu2024survey, li2024llms} and to work on judge bias, benchmark alignment, and general judge design~\cite{shi2025judging, wataoka2024self, chen2024humans, wang2025assessing, tan2024judgebench, jiang2025codejudgebench}, it adds a measurement view: existing pipelines produce judgments, whereas here the judgments of four judges are checked as measurements, scored against objective contest verdicts, probed with controlled defect injection, and reported for where they hold and where they do not. Perturbation-based benchmarks of code judges~\cite{wang2025axiom} likewise report weaknesses in consistency and refinement-effort judgment. The panel examined here combines the four judges by unweighted percentile rank; it gains about 0.07 ROC-AUC over the mean member while remaining below its strongest member (interval excluding zero; Section~\ref{integrated}), a baseline for dynamically selected teams such as SE-Jury~\cite{zhou2025se}. Like Code-DiTing~\cite{yang2025code}, the judges here assess code against the problem statement without reference solutions or test cases, and their outputs are scored against objective execution verdicts, which links reference-free judging to test-based ground truth.

\section{Motivation and Task Design}
\label{TD}
\subsection{Motivation}
Programming education is caught between two realities: AI systems now generate code and explanations at scale, yet it remains unclear which competencies learners must still internalize to remain reliable, independent problem solvers. Competitive programming is a stringent test bed: success requires (i) core competencies (algorithms, data structures, control flow) and (ii) higher-order skills (critical evaluation, problem formulation, ethical judgment, domain intuition) to verify, correct, and extend AI suggestions under time pressure. Its co-creative traces contain partial attempts, revisions, and intermediate reasoning, and a pass/fail signal does not show how a solution is reached. This motivates a task setting and measurement perspective for reproducible analysis of co-creation behavior and the limits of AI assistance, anchored to objective execution outcomes, which also serve as the ground truth against which the judges are measured (Section~\ref{TD:whyjudge}).

\subsubsection{Study Setting: Parallel Contest vs.\ AI-Permitted Co-Creation Track}
\label{TD:setting}
The study uses the programming division of the PCK Finals at the University of Aizu~\cite{PCK}. AI/LLM use is forbidden in the official track; in a parallel AI-permitted co-creation track, a different group of participants attempts the same final-round problems, released live, with any AI tools, in the same four-hour window and under the same rules apart from AI permission. Prompting, verification, and revision are therefore observed under genuine contest time pressure. The contest committee authors the problems without AI assistance, and they appear in the contest judge system only when the contest begins, so neither the participants' tools nor the judges used here had access to them beforehand. The submitted programs and verdicts remain in that system and are not public, so the labeled pairs against which the judges are scored have no public counterpart. The setting supplies an objective verdict for every attempt and the participants' interaction logs alongside the submissions; it does not supply the breadth of everyday development, since the participants, problems, and time limit are those of a single contest.

\subsubsection{Why an LLM Judge Alongside an Execution Verdict}
\label{TD:whyjudge}
The contest verdict is objective but coarse: it records whether an attempt was accepted and, if not, only a broad failure category (wrong answer, time limit, or compile error), not how close the attempt was, so a near-correct attempt and a fundamentally flawed one can receive the same verdict. A graded, criterion-level assessment makes intermediate progress measurable, which trajectory-level analysis requires. The judge does not replace the verdict; the verdict makes the judge measurable: every attempt carries an objective label, so judge quality is scored against ground truth, not agreement with annotators, and, given the provenance above (Section~\ref{TD:setting}), without entanglement with benchmark memorization. A judge can also be probed where a verdict cannot: a defect can be injected into a program and the response of the score observed, as the perturbation study does (Section~\ref{sec:perturbation_design}), which makes an LLM judge auditable in a way a pass or fail signal is not.

\subsubsection{Reliability and Validity}
\label{TD:relval}
Following standard measurement usage, \emph{reliability} refers to consistency: the same instrument produces the same measurement across repeated application, and different instruments applied to the same object agree. \emph{Validity} refers to whether a measurement tracks the property it is meant to capture. A pipeline is \emph{reliability-aware} when it reports both, together with the conditions under which each degrades, and not a score alone. Here, reliability is examined through inter-judge agreement and through repeated judgments of identical code; validity through discrimination against the contest verdict and through the perturbation study (Section~\ref{sec:perturbation_results}); and the conditions of degradation through the effect of input specification, the resolution of each judge's output scale, and calibration error (Section~\ref{results_judge}).

\subsection{Task Design and Data Representation}
During the AI-permitted co-creation track, participant activity is captured through submission logs and interaction traces, yielding chronological records for each participant-problem pair. Each submission attempt records a participant $u$, a problem $p$, its index within that pair, the submitted code, the prompt-side interaction text associated with it, a timestamp, and the external execution verdict returned by the contest judge system. Every judge request carries the problem statement and the submitted code; three input conditions, \textsc{blind}, \textsc{prefix12k} and \textsc{nonblind}, differ only in the participant context added, from none to a problem-specific context cut where the submitted code first appears in the logs (Section~\ref{sec:approach_data}). A binary success label is derived from the verdict, which anchors the study to an objective outcome while still permitting analysis of intermediate progress. Attempts are grouped into trajectories by $(u,p)$, with temporal order preserved to support turn-level characterization of revision behavior and time-to-success.

\subsubsection{Measurement Targets}
The task needs measurements beyond final correctness that stay compatible with contest outcomes. Two targets are used: the predicted acceptance probability of an attempt, given the problem statement, its code and, under \textsc{prefix12k} and \textsc{nonblind}, participant context; and criterion-level assessments of algorithmic adequacy and robustness/constraint satisfaction, which separate a near-correct attempt from a fundamentally flawed one with the same verdict. Metrics are defined in Section~\ref{eval}.

\subsubsection{Research Questions and Evaluation Questions}
The study is organized around three pairs of questions. Each evaluation question (EQ) specifies what a valid measurement pipeline must demonstrate, and the research question (RQ) that follows it is answered only through measurements that the EQ shows to be valid.

\noindent \textbf{EQ1 (Trajectory validity):} Do attempt-level signals support coherent trajectory analyses, including Success@Turn (success by turn index $k$) and time-to-success under right-censoring, without relying solely on the final attempt?

\noindent \textbf{RQ1 (Co-creation dynamics):} How do participants progress across successive attempts when co-creating with AI under competitive constraints, and what turn-level patterns characterize both struggle and eventual success?

\medskip
\noindent \textbf{EQ2 (Churn \& similarity validity):} Can revision magnitude be quantified robustly and comparably across participants/problems using turn-to-turn churn and code-aware similarity measures (e.g., NED and CodeBLEU), and do these measures relate meaningfully to progress signals such as changes in the judges' rank consensus?
 
\noindent \textbf{RQ2 (Revision behavior):} How are revision behaviors related to progress across successive attempts, and can code-side changes distinguish incremental fixing from larger rewrites with limited gains?

\medskip
\noindent \textbf{EQ3 (Measurement quality):} Under grouped cross-validation (CV) over all attempts, do judge-produced outputs exhibit (i) discrimination against the contest verdict (ROC-AUC/PR-AUC), (ii) output scales that support stable thresholded decisions, with thresholds selected out of fold (MCC), (iii) probabilistic calibration (LogLoss/Brier/ECE), (iv) inter-judge agreement (Cohen's/Fleiss' $\kappa$), and (v) a response to injected defects that exceeds the response to behavior-preserving edits?
 
\noindent \textbf{RQ3 (Judges as measurement instruments):} Can multiple LLM judges provide auditable and comparable assessments of co-creative attempts, and how stable are these measurements across judges?

\medskip
The judge-dependent measurements in RQ1 and RQ2 (the turn-wise rank consensus of the judges and judged improvement) also rest on EQ3, whereas Success@Turn, time-to-success, revision churn, and code-aware similarity are computed from submissions and verdicts alone and do not depend on any judge output. The results therefore present the evaluation questions first, and each research-question answer names the evaluation result it depends on.

\section{Methodology}
\label{PA}
\begin{figure*}[t]
\centering
\includegraphics[width=5.8in]{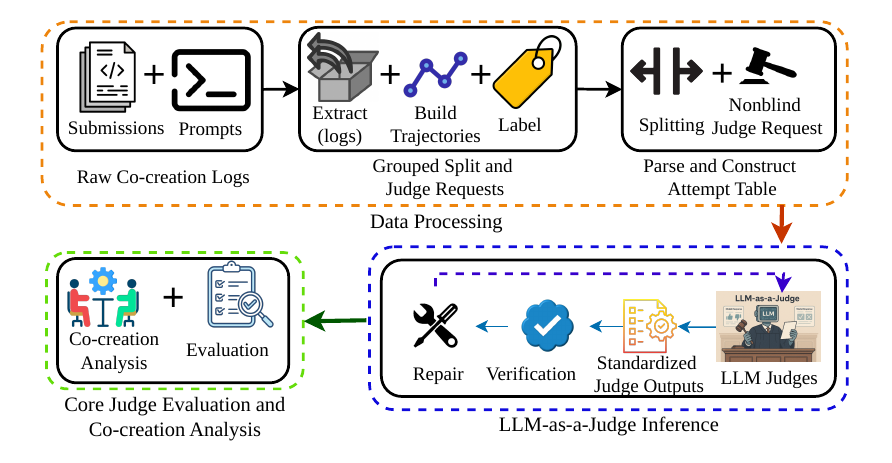}
\caption{Overview of the \emph{LLM-as-a-Judge} and \emph{human-AI co-creation} workflow}
\label{approach_workflow}
\end{figure*}

The pipeline (\figboxref{approach_workflow},~\algboxref{workflow}) has three stages: (i) data processing and attempt-level dataset construction from submission artifacts and prompt logs, (ii) API-based judge inference under a unified schema with verification and bounded repair, and (iii) consolidation, judge evaluation, and trajectory-level co-creation analysis.~\tabboxref{tab:notation} collects the notation used in Sections~\ref{TD} to~\ref{Dis}.

\begin{algorithm}[h]
\caption{\emph{LLM-as-a-Judge Workflow} for \emph{AI Co-Creation}}
\label{workflow}
\begin{algorithmic}[1]
\REQUIRE Raw submissions $\mathcal{S}$, prompt logs $\mathcal{P}$, judges $\mathcal{J}$, input conditions $\{\textsc{blind},\textsc{prefix12k},\textsc{nonblind}\}$, fold count $K$
\ENSURE Verified judge outputs $\{O_j\}_{j\in\mathcal{J}}$, merged prediction table $\mathcal{T}$, and evaluation artifacts
\STATE Parse $\mathcal{S}$ and $\mathcal{P}$; construct attempt table $\mathcal{A}=\{(a,u,p,k,c,h,\tau,v,y)\}$ sorted by $(u,p,\tau)$
\STATE For each $a\in\mathcal{A}$, locate $\mathrm{anchor}(a)$, the first occurrence of $c_a$ in $u$'s logs when present, and construct $x_a^{(\mathrm{cond})}$ for $\mathrm{cond}\in\{$ \textsc{blind}, \textsc{prefix12k}, \textsc{nonblind}$\}$
\STATE Split trajectories by group key $(u,p)$ into $K$ folds; serialize canonical judge requests $\mathcal{R}=\{(a,\mathrm{cond},x_a^{(\mathrm{cond})})\}$ for every attempt and condition
\STATE \textbf{for each} judge $j\in\mathcal{J}$ \textbf{in parallel do}
\STATE \hspace{4mm} infer outputs under a strict JSON schema with checkpointing
\STATE \hspace{4mm} verify every field; \textbf{while} an \texttt{attempt\_id} fails validation \textbf{and} its retry budget remains \textbf{do} reissue the identical request under bounded retry with backoff
\STATE \hspace{4mm} record any \texttt{attempt\_id} that exhausts its retry budget as a failure, without reducing its input
\STATE \textbf{end for}
\STATE Merge all verified outputs with $\mathcal{A}$ by \texttt{attempt\_id} to obtain $\mathcal{T}$
\STATE \textbf{for each} fold $k\in\{1,\dots,K\}$ \textbf{do} select threshold $t_j^\star$ on the other $K-1$ folds and apply it to fold $k$, for every judge $j$
\STATE Compute cross-judge agreement and trajectory-level dynamics; produce final evaluation artifacts
\end{algorithmic}
\end{algorithm}

\begin{table}[h]
\centering
\caption{Notation used in Sections III--VII}
\setlength{\arrayrulewidth}{0.2mm}
\setlength{\tabcolsep}{1pt}
\renewcommand{\arraystretch}{1}
\label{tab:notation}
\small
\begin{tabular}{@{}l||l@{}}
\hline
\textbf{Symbol} & \textbf{Meaning} \\ \hline
\hline
$u$, $p$ & participant identifier, problem identifier \\ \hline
$a$ & \shortstack{attempt identifier (\texttt{attempt\_id}); primary key \\for every judge and analysis table} \\ \hline
$(u,p,k)$ & \shortstack{trajectory and turn index of attempt $a$: the $k$-th \\attempt by $u$ on $p$} \\ \hline
$c_a, h_a, \tau_a, v_a$ & \shortstack{submitted code, prompt-side text, timestamp, \\and verdict of attempt $a$} \\ \hline
$y_a$ & ground-truth acceptance label, $y_a=\mathbb{I}[v_a=\mathrm{AC}]$ \\ \hline
$\mathrm{cond}$ & \shortstack{input condition, \\ $\mathrm{cond}\in\{\textsc{blind},\textsc{prefix12k},\textsc{nonblind}\}$} \\ \hline
$x_a^{(\mathrm{cond})}$ & judge request for attempt $a$ under condition $\mathrm{cond}$ \\ \hline
$\mathcal{J}$ & \shortstack{set of judges (OpenAI, DeepSeek, Gemini, \\Claude)} \\ \hline
$p_a^{(j)}$ & \shortstack{judge $j$'s predicted acceptance probability for \\attempt $a$} \\ \hline
$s_{\mathrm{algo},a}^{(j)}, s_{\mathrm{robust},a}^{(j)}$ & \shortstack{judge $j$'s ordinal scores for algorithmic \\adequacy, robustness} \\ \hline
$t_j^\star$ & \shortstack{judge $j$'s accept/reject threshold, selected out \\of fold} \\ \hline
$\hat y_a^{(j)}$ & \shortstack{judge $j$'s predicted acceptance label, \\ $\hat y_a^{(j)}=\mathbb{I}[p_a^{(j)}\ge t_j^\star]$} \\ \hline
$\bar q_{u,p,k}$ & \shortstack{rank consensus of $\mathcal{J}$ at turn $k$ of $(u,p)$ \\(mean percentile rank of $p$)} \\ \hline
$\Delta \bar q_{u,p,k}$ & turn-wise change in $\bar q_{u,p,k}$ (judged improvement) \\ \hline
$\mathcal{G}$ & set of participant-problem trajectories \\ \hline
$K_{u,p}$, $T^\star_{u,p}$ & \shortstack{last observed turn, and latent first-success turn, \\of $(u,p)$} \\ \hline
\hline
\end{tabular}
\end{table}

\subsection{Data Processing and Request Construction}
\label{sec:approach_data}
\subsubsection{Raw artifacts and canonical parsing}
Each participant record consists of (i) submission artifacts from the competition judge system, containing the submitted code and verdict, and (ii) prompt logs of the participant's interaction with AI tools. Submission artifacts are parsed from text-like formats (\texttt{.html}, \texttt{.htm}, \texttt{.md}, \texttt{.txt}) to avoid optical character recognition (OCR) noise. For each submission, the participant \(u\), problem \(p\), timestamp \(\tau\), source code \(c\), verdict \(v\), and language \(\ell\) are extracted, the verdict and the timestamp each from its own named field, not from the visible page text (Section~\ref{TD:dataverif}, V1 and V3). For HTML pages, scripts and styles are removed first, and the submitted code is taken from the longest \texttt{<pre>} block, with \texttt{<code>} as a fallback.

\subsubsection{Attempt-level dataset construction}
Parsed submissions are sorted by \((u,p,\tau)\). Within each participant-problem group, a trajectory turn index \(k \in \{1,2,\dots\}\) and a stable global identifier \(a\) (\texttt{attempt\_id}) are assigned: for instance, the third submission by participant $u$ on problem $p$ is turn $k=3$ of trajectory $(u,p)$ and receives its own $a$, which is used as the primary key for all downstream joins. Each attempt is represented as
\vspace{-0.95em}
\begin{equation}
a=\left(u,p,k,c_{a},h_{a},\tau_{a},v_{a}\right),
\end{equation}
and the ground-truth acceptance label is
\vspace{-0.5em}
\begin{equation}
y_{a}=\mathbb{I}\!\left[v_{a}=\mathrm{AC}\right],
\label{eq:gt}
\end{equation}

where \(\mathbb{I}[\cdot]\) denotes the indicator function and \(h_a\) is the prompt-side interaction text associated with attempt $a$. The participant context actually sent to a judge is constructed separately, as described next.

\subsubsection{Anchor-Based Context and Input Conditions}
Prompt logs are parsed into textual records and ordered deterministically (e.g., by filename and/or stable file metadata). Because prompt timestamps are not uniformly reliable across all participants and formats, temporal truncation cannot be implemented from timestamps alone. Instead, for each attempt $a$ whose code $c_a$ can be located verbatim in participant $u$'s logs, an anchor point is defined at its first appearance, and the context supplied to a judge is drawn only from log content strictly preceding that point, restricted to turns that concern problem $p$. Every judge request carries the problem statement and the submitted code; the three input conditions differ only in what participant context, if any, is added: $(i)$ \textsc{blind}: no participant context. $(ii)$ \textsc{prefix12k}: the first 12,000 characters of the participant's logs, aggregated across all of $u$'s problems in a fixed deterministic order, independent of $p$ and of any anchor. $(iii)$ \textsc{nonblind}: the anchored context defined above, capped at 12,000 characters, restricted to problem $p$.
 
Context provenance for the \textsc{nonblind} condition is reported in three tiers. For 369 of the 517 attempts (71.4\%), the code can be located in the logs, and the context is \emph{anchored}, meaning it is proven to precede the submission it accompanies. For 102 (19.7\%), the correct problem's log content can be identified without locating the code itself, and the context is \emph{segmented} to that problem without an anchor. For the remaining 46 (8.9\%), neither is possible, and no context is sent, since context about the wrong problem would be worse than none. Anchored and segmented attempts together, 471 (91.1\%), are termed evidence-backed. \textbf{Participant identity is not disclosed to a judge under any condition: all context is scrubbed of personal information before a request is issued.}

\subsubsection{Grouped splitting and request serialization}
To prevent leakage across repeated attempts of the same participant on the same problem, grouped CV folds attempts by trajectory key \(g=(u,p)\), so that no trajectory's threshold is selected on its own attempts (Section~\ref{sec:approach_analysis}). For each attempt \(a\) and input condition \(\mathrm{cond}\), a canonical request is serialized as

\vspace{-0.5em}
\begin{equation}
\label{eq:req}
x_a^{(\mathrm{cond})}=
\left\{
\begin{aligned}
&\texttt{problem\_id},\\
&\texttt{problem\_statement},\\
&\texttt{language},\ \texttt{source\_code},\\
&\texttt{prompt\_text}^{(\mathrm{cond})}
\end{aligned}
\right\},
\end{equation}
 
where \texttt{problem\_statement} is the full statement, including input constraints, expected behavior, and examples, and \texttt{prompt\_text}$^{(\mathrm{cond})}$ is the participant context defined above for condition $\mathrm{cond}$ (empty under \textsc{blind}). All requests are stored in a model-agnostic JSONL file shared by all judges. Ground-truth labels are excluded from these requests to avoid label leakage.

\subsection{LLM-as-a-Judge Inference and Verification}
\label{sec:approach_judge}
\subsubsection{Unified judge schema}
Multiple judges \(j\in\mathcal{J}\) (e.g., OpenAI, DeepSeek, Gemini, and Claude) are evaluated through API-based inference. The same input \(x_a^{(\mathrm{cond})}\) is provided to each judge, for each condition in turn, and a strict JSON object of the following form is required:
\vspace{-0.2em}
\begin{equation}
\hat{z}^{(j)}_a=
\left(
p^{(j)}_a,\;
s^{(j)}_{\mathrm{algo},a},\;
s^{(j)}_{\mathrm{robust},a},\;
r^{(j)}_a
\right),
\label{eq:judge_output}
\end{equation}
 
where \(p^{(j)}_a\in[0,1]\) denotes the predicted acceptance probability, \(s^{(j)}_{\mathrm{algo},a}\in\{1,\dots,5\}\) denotes an ordinal score for algorithmic adequacy, \(s^{(j)}_{\mathrm{robust},a}\in\{1,\dots,5\}\) denotes an ordinal score for robustness and constraint handling, and \(r^{(j)}_a\) denotes a short rationale. The discrete ranges \(\{1,\dots,5\}\) represent bounded rubric levels, where lower values indicate poorer performance and higher values indicate stronger satisfaction of the corresponding criterion. All judges are executed under the same schema, on a single shared request-and-validation path rather than a runner written per judge, so that the resulting outputs are directly comparable across models.

\subsubsection{Verification and bounded repair}
Inference is executed with checkpointing so that successful attempts are not recomputed on reruns. Because large-scale API calls can produce transient failures (e.g., rate limits, timeouts, quota limits, or truncated responses), bounded retries with backoff are applied to failed \texttt{attempt\_id}s at the transport level, independently of the validation check below. A response is accepted only if all of the following hold:
\begin{equation}
\begin{aligned}
&\text{(a) the response parses as the schema of Eq.~\eqref{eq:judge_output};}\\
&\text{(b) every field of \(\hat z^{(j)}_a\) is present;}\\
&\text{(c) } p^{(j)}_a \in [0,1];\\
&\text{(d) } s^{(j)}_{\mathrm{algo},a} \in \{1,\dots,5\} \text{ and } s^{(j)}_{\mathrm{robust},a} \in \{1,\dots,5\}.
\end{aligned}
\label{eq:verify}
\end{equation}
 
A judge output table \(O_j\) is accepted only once it contains exactly one row satisfying (a)-(d) for every \(a\in\mathcal{A}\), with no attempt missing and none duplicated. When a response fails any condition in Eq.~\eqref{eq:verify}, repair applies: the identical request is reissued, unchanged, under bounded retry with backoff, for that \texttt{attempt\_id} alone. The input is not reduced on repair, since a row judged on a shorter input than every other row would no longer be comparable to them; an \texttt{attempt\_id} that exhausts its retry budget is recorded as a failure instead. A repaired response overwrites the earlier failure state, preserving a one-to-one mapping between \texttt{attempt\_id} and the final judge output. Only verified outputs are passed to downstream evaluation. In the evaluation run, 11 of the 6,204 responses failed validation on first receipt, all through truncated or empty output, and each returned a valid response on reissue.

\subsection{Consolidation and Analysis Protocol}
\label{sec:approach_analysis}
\subsubsection{Unified prediction tables}
All verified judge outputs are merged with the attempt table by \texttt{attempt\_id}. Long-format and wide-format prediction tables are then retained for grouped analyses, plotting, and cross-judge comparisons.

\subsubsection{Judge evaluation protocol}
For each judge and input condition, probabilistic predictions are evaluated by grouped $K$-fold CV over all 517 attempts, with folds formed on the trajectory key $(u,p)$ so that no trajectory is split across folds and every attempt is scored out of fold exactly once. Within each fold, the accept/reject threshold \(t_j^\star\) is selected on the other $K-1$ folds by maximizing the Matthews Correlation Coefficient (MCC) and applied, unchanged, to the held-out fold, giving the predicted acceptance label \(\hat y_a^{(j)}=\mathbb{I}[p_a^{(j)}\ge t_j^\star]\). Discrimination, calibration, and thresholded decision metrics are then computed by pooling the out-of-fold \(p_a^{(j)}\) and \(\hat y_a^{(j)}\) across folds. Cross-judge agreement instead uses each judge's continuous output cut at its own median, with rank agreement (Spearman's $\rho$) reported alongside it. Confidence intervals are obtained by bootstrap resampling the trajectories $(u,p)$, which are the unit of dependence in this corpus; a problem-clustered re-estimation, resampling the 13 problems instead, is reported alongside it as a robustness check.

\subsubsection{Trajectory construction and co-creation dynamics}
For each trajectory \((u,p)\), the judges' rank consensus at turn $k$ is computed as
\vspace{-0.8em}
\begin{equation}
\bar{q}_{u,p,k}
=
\frac{1}{|\mathcal{J}|}
\sum_{j\in\mathcal{J}} F_j\!\left(p^{(j)}_{u,p,k}\right),
\end{equation}
where $F_j$ gives the percentile rank of a value among the outputs of judge $j$ on all 517 attempts under \textsc{nonblind}, ties sharing their average rank. Ranking each judge first lets judges of very different output resolution (Section~\ref{sec:scale_validity}) count equally. The result is the score of the unweighted percentile-rank panel evaluated in Section~\ref{results_judge}, and it orders the attempts almost exactly as a plain mean of the probabilities would (Spearman 0.974 across the 517 attempts). The corresponding turn-wise progress, the judged improvement, is defined as
\vspace{-0.5em}
\begin{equation}
\Delta \bar{q}_{u,p,k}
=
\bar{q}_{u,p,k}-\bar{q}_{u,p,k-1},
\qquad k\ge 2.
\end{equation}
 
These quantities are used to characterize struggle curves, recovery patterns, and stagnation over iterative attempts.

\subsubsection{Time-to-success and similarity analyses}
Let \(K_{u,p}\) denote the last observed turn index for trajectory \((u,p)\). The latent first-success turn is
\vspace{-0.7em}
\begin{equation}
T^{\star}_{u,p}
=
\min\{k:\ y_{u,p,k}=1\},
\end{equation}
with \(T^{\star}_{u,p}=+\infty\) if the trajectory is never solved. For survival analysis~\cite{kaplan1958nonparametric}, the observed time and censoring indicator are
\vspace{-0.5em}
\begin{equation}
T_{u,p}=\min(T^{\star}_{u,p},K_{u,p}),
\qquad
\delta_{u,p}=\mathbb{I}[T^{\star}_{u,p}\le K_{u,p}],
\end{equation}
thus treating unsolved trajectories as right-censored. Success by turn index \(k\) is reported as
\vspace{-0.6em}
\begin{equation}
S(k)
=
\frac{1}{|\mathcal{G}|}
\sum_{(u,p)\in\mathcal{G}}
\mathbb{I}[T^{\star}_{u,p}\le k],
\end{equation}

where \(\mathcal{G}\) is the set of all participant-problem trajectories. Prompt churn, code churn, and code-aware similarity between consecutive turns quantify revision behavior; the analysis uses code churn and CodeBLEU similarity (Section~\ref{eval}).

\subsection{Data Verification}
\label{TD:dataverif}
Because this revision re-estimates every reported value, the data pipeline was verified end to end before any analysis was rebuilt. The analysis code (metric definitions, MCC protocol, calibration, agreement statistics, and survival analysis) behaved as specified; the verification found three issues in its input, V1, V3, and V4, with V2 a consequence of V1 and V5 a reproducibility check on the parsing layer.

\textbf{V1. Submission timestamps.} Each saved submission page carries two clocks: a header clock recording when the page was rendered, and a named \texttt{Submit} field recording when the code was submitted. Timestamps are read from the \texttt{Submit} field, with the header clock retained separately as \texttt{page\_saved\_ts} so the two are not conflated. The corrected timestamps span 13:49:28 to 17:44:50 on the contest date, inside the contest's official 13:45-17:45 window, and order 72 of 73 multi-attempt trajectories identically to the independent filename submission order.

\textbf{V2. Trajectory ordering.} This follows from V1: pages had been saved newest-first from a submissions list, so the corrected clock reverses the order within each multi-attempt trajectory, which is why Success@Turn now rises across turns rather than beginning near its asymptote.

\textbf{V3. Verdict extraction.} Verdicts are read from the named \texttt{Result} field, no longer by scanning the visible page text, since submitted source code is part of that text and, for one of the thirteen problems, solutions commonly declare a variable named \texttt{accepted}, which a whole-page scan can match in place of the verdict itself. All 517 pages were audited and contain exactly four distinct \texttt{Result} strings, all mapped; every disagreement between the two extraction methods falls on that single problem and no other.

\textbf{V4. Join key.} The \texttt{attempt\_id} used to associate judge outputs with attempts had been assigned by position after sorting; since V1 changes the sort order, the correspondence was rebuilt through the source file path, adding a stable natural key, \texttt{attempt\_uid}. The submitted code each judge was sent was compared against the code of the attempt its prediction is now attached to, and matches on 517 of 517 rows.

\textbf{V5. Pipeline reproducibility.} The parsing layer was rebuilt independently and its output compared against the frozen archive of the earlier analysis: timestamps, problem identifiers, submission identifiers, language, and code length match on every row, and the full pipeline runs end to end from a clean environment.

\textbf{Effect on the reported corpus.} Reading verdicts from the named field (V3) changes the label on 35 attempts recorded as accepted in the data as originally analyzed, all on that single problem: accepted attempts fall from 193 to 158 (with 192 wrong-answer, 147 time-limit-exceeded, and 20 compile-error;~\tabboxref{data_overview}), and prevalence from 0.3733 to 0.3056. The same labels move the convergence observations from 329 to 280 and the design-phase partition, stratified on whether a trajectory is ever accepted, from 148/18/18 trajectories and 373/108/36 attempts to 146/19/19 and 401/66/50; V2 moves Success@Turn at turn 1 from 0.8533 to 0.6033. This revision reports nothing from either partition, since every judge figure comes from CV over all 517 attempts (Section~\ref{sec:approach_analysis}).

\section{Experiments}
\label{Ex}
\subsection{Dataset Overview}
\label{sec:dataset_overview}
\tabboxref{data_overview} summarizes the dataset. Each trajectory \((u,p)\) is a chronological sequence of attempts, and each attempt holds the submitted code, prompt-side context, timestamp, verdict, and acceptance label. The interaction logs range from conversational use of chat assistants to full delegation, in which one participant configured autonomous coding agents to read the problems, write the solutions, and submit them. Reading a specification, writing code, submitting it, receiving an objective pass or fail from execution, diagnosing the failure, and revising is the inner loop of software development that continuous integration automates; this setting differs from everyday practice in scale and duration, not in the kind of work. A grouped train/validation/test split decided, before any judge prompt was written, which trajectories could be inspected during prompt and schema design (Section~\ref{TD:dataverif}); judge evaluation uses grouped CV over all attempts (Section~\ref{sec:approach_analysis}).

\begin{table}[t]
\caption{Overview of the co-creation dataset}
\setlength{\arrayrulewidth}{0.2mm}
\setlength{\tabcolsep}{9.5pt}
\renewcommand{\arraystretch}{1}
\label{data_overview}
\begin{tabular}{l||l}
\hline \hline
\textbf{Statistic}                  & \textbf{Value}                 \\ \hline \hline
Participants ; Problems & 15 ; 13 \\ \hline 
Participant-problem trajectories \((u,p)\) & 184 \\ \hline 
Submission attempts (rows) ; Prompt logs & 517 ; 83 \\ \hline 
Accepted (AC) ; Wrong answer (WA) & 158 ; 192 \\ \hline
Time limit exceeded (TLE) ; Compile error (CE) & 147 ; 20 \\ \hline
Design-phase train / val / test trajectories & 146 / 19 / 19 \\ \hline
Average attempts per trajectory & 2.81 \\ \hline
Median attempts per trajectory & 1 \\ \hline 
Maximum attempts in a trajectory & 30 \\ \hline \hline
\end{tabular}
\end{table}

\subsection{Evaluation Metrics}
\label{eval}
The pipeline is evaluated as a measurement instrument and through co-creation dynamics. Each metric is reported with the decision it serves: discrimination when a judge ranks or triages attempts, calibration when a threshold is applied to its output, and defect sensitivity when a judge assesses quality instead of predicting an outcome;~\tabboxref{metric_summary} lists each metric, its intended use, and its role.

\begin{table}[h]
\caption{Summary of evaluation metrics, their intended use, and their roles}
\setlength{\arrayrulewidth}{0.2mm}
\setlength{\tabcolsep}{2.2pt}
\renewcommand{\arraystretch}{1.1}
\label{metric_summary}
\begin{tabular}{l||c||l}
\hline \hline
\textbf{Metric} & \textbf{Decisive for} & \textbf{Role in the study} \\ \hline \hline
ROC-AUC & \shortstack{Ranking, \\triage} & \shortstack{Ranking ability for accepted vs. \\non-accepted attempts} \\ \hline
PR-AUC & \shortstack{Ranking, \\triage} & \shortstack{Precision-recall behavior under class \\imbalance} \\ \hline
LogLoss & \shortstack{Threshold \\transfer} & \shortstack{Penalization of inaccurate and \\overconfident probability predictions} \\ \hline
Brier score & \shortstack{Threshold \\transfer} & \shortstack{Mean squared error of probabilistic \\predictions} \\ \hline
ECE & \shortstack{Threshold \\transfer} & \shortstack{Agreement between predicted probability \\and observed acceptance rate} \\ \hline
MCC & \shortstack{Decision \\quality} & \shortstack{Balanced decision quality by \\ out-of-fold-selected threshold} \\ \hline
Cohen's $\kappa$ & \shortstack{Cross-judge \\reliability} & \shortstack{Pairwise agreement between judges} \\ \hline
Fleiss' $\kappa$ & \shortstack{Cross-judge \\reliability} & \shortstack{Multi-rater agreement across all judges} \\ \hline
Spearman's $\rho$ & \shortstack{Cross-judge \\reliability} & \shortstack{Rank agreement between judge pairs} \\ \hline
Struggle curves & -- & \shortstack{Turn-wise rank consensus of the judges \\dynamics over trajectories} \\ \hline
Success@Turn & -- & \shortstack{Fraction of trajectories solved by or \\before turn index \(k\)} \\ \hline
\shortstack{Kaplan-Meier \\survival} & -- & \shortstack{Time-to-success with right-censoring \\at the last observed turn} \\ \hline
\shortstack{Prompt/code \\NED churn} & -- & \shortstack{Surface-form revision magnitude \\between consecutive turns} \\ \hline
\shortstack{CodeBLEU \\churn} & -- & \shortstack{Code-aware change between consecutive \\turns} \\ \hline
\shortstack{CodeBLEU \\convergence} & -- & \shortstack{Similarity to the first accepted \\solution within a solved trajectory} \\ \hline
\shortstack{Arm \\discrimination} & \shortstack{Defect \\sensitivity} & \shortstack{Separation between injected-defect \\and matched control edits} \\ \hline
\shortstack{Detection \\specificity} & \shortstack{Defect \\sensitivity} & \shortstack{Genuine detections against false alarms \\on behavior-preserving edits} \\ \hline
\shortstack{Severity \\gradient} & \shortstack{Defect \\sensitivity} & \shortstack{Whether the score tracks how visible a \\defect is, not merely that something \\changed} \\ \hline \hline
\end{tabular}
\end{table}

\subsubsection{Judge quality metrics}
\paragraph{ROC-AUC and PR-AUC (discrimination metrics; decisive for ranking and triage)}
Discrimination metrics quantify how well accepted attempts are ranked above non-accepted attempts using predicted probabilities \(p_a\) without fixing a decision threshold~\cite{fawcett2006introduction}. ROC-AUC measures the probability that a randomly chosen accepted attempt receives a higher score than a randomly chosen non-accepted attempt. PR-AUC (reported as Average Precision) summarizes the precision-recall trade-off and is especially informative under class imbalance. Higher values indicate better discrimination.

\paragraph{LogLoss ($\mathcal{L}$) (probabilistic scoring; decisive when a threshold is applied downstream)}
LogLoss evaluates the quality of probabilistic predictions on a set of \(N\) attempts and penalizes overconfident errors strongly~\cite{gneiting2007strictly}:
\vspace{-0.8em}
\begin{equation}
\mathcal{L}
=
-\frac{1}{N}\sum_{a=1}^{N}
\left[
y_a\log p_a + (1-y_a)\log(1-p_a)
\right].
\end{equation}
Lower values indicate better probabilistic accuracy.

\paragraph{Brier score ($B$) (probabilistic scoring; decisive when a threshold is applied downstream)}
The Brier score measures mean squared error of probabilistic predictions~\cite{glenn1950verification}:
\vspace{-0.8em}
\begin{equation}
\mathrm{B}
=
\frac{1}{N}\sum_{a=1}^{N}(p_a-y_a)^2.
\end{equation}
where lower values indicate better probabilistic accuracy.

\paragraph{Expected Calibration Error (ECE) (calibration metric; decisive when a threshold is applied downstream)}
Calibration metrics quantify how well predicted probabilities match empirical frequencies. ECE is computed by partitioning the predicted acceptance probabilities into \(B\) equal-width bins and comparing, in each bin, the observed acceptance rate with the mean predicted probability~\cite{guo2017calibration}:
\vspace{-0.1em}
\begin{equation}
\mathrm{ECE}
=
\sum_{b=1}^{B}\frac{n_b}{N}
\left|
\mathrm{obs}(b)-\mathrm{pred}(b)
\right|,
\qquad B=15,
\end{equation}
where \(n_b\) is the number of predictions in bin \(b\).

\paragraph{Matthews Correlation Coefficient (MCC) (thresholded decision quality)}
Thresholded decision metrics evaluate binary accept/reject decisions after converting probabilities to labels. Decisions are obtained as
\begin{equation}
\hat{y}_a(t)=\mathbb{I}[p_a\ge t].
\end{equation}
The threshold is selected, for each CV fold, on the other folds by maximizing MCC and applied, unchanged, to the held-out fold:
\begin{equation}
t^\star=\arg\max_{t\in[0,1]}\mathrm{MCC}_{\mathrm{other\ folds}}(t).
\end{equation}
MCC is defined as~\cite{matthews1975comparison}
\begin{equation}
\mathrm{MCC}
=
\frac{TP\cdot TN - FP\cdot FN}
{\sqrt{(TP+FP)(TP+FN)(TN+FP)(TN+FN)}},
\end{equation}
where \(TP\), \(TN\), \(FP\), and \(FN\) denote true positives, true negatives, false positives, and false negatives, respectively.

\paragraph{Cohen's $\kappa$ and Fleiss' $\kappa$ (agreement / reliability metrics)}
Agreement metrics treat LLM judges as raters and quantify reliability beyond chance agreement. Cohen's \(\kappa\) is used for pairwise agreement~\cite{cohen1960coefficient}:
\begin{equation}
\kappa
=
\frac{p_o-p_e}{1-p_e},
\end{equation}
where \(p_o\) is observed agreement and \(p_e\) is agreement expected by chance. Fleiss' \(\kappa\) extends this notion to the multi-rater setting~\cite{fleiss1971measuring}. Higher \(\kappa\) indicates stronger agreement. Rank agreement between two judges is measured with Spearman's $\rho$ on their continuous outputs, reported alongside $\kappa$ so that agreement on ordering can be told apart from agreement on a cut.

\medskip
\subsubsection{Trajectory and co-creation metrics}
\paragraph{Struggle curves, Success@Turn, and time-to-success (trajectory dynamics)}
These signals, the judges' rank consensus at a turn, its turn-wise change, success by turn index, and Kaplan-Meier survival, are defined once, in Section~\ref{sec:approach_analysis}; success by turn index and survival need only the verdicts, while the other two rest on the judges' \textsc{nonblind} outputs.

\paragraph{Normalized Edit Distance (NED) (string-based churn)}
String-based churn is quantified using NED based on Levenshtein distance~\cite{levenshtein1966binary}:
\vspace{-0.6em}
\begin{equation}
\mathrm{NED}(s,t)
=
\frac{\mathrm{LD}(s,t)}
{\max(1,|s|,|t|)},
\end{equation}
where \(\mathrm{LD}(\cdot,\cdot)\) is Levenshtein distance. For prompt-side text \(h_{u,p,k}\) and code \(c_{u,p,k}\), churn between consecutive observed turns is defined as
\begin{equation}
\begin{aligned}
\mathrm{Churn}^{\mathrm{prompt}}_{u,p,k}
&=
\mathrm{NED}(h_{u,p,k-1},h_{u,p,k}),\\
\mathrm{Churn}^{\mathrm{code}}_{u,p,k}
&=
\mathrm{NED}(c_{u,p,k-1},c_{u,p,k}),
\qquad k\ge 2.
\end{aligned}
\end{equation}

\paragraph{CodeBLEU-based churn and convergence (code-aware similarity)}
To reduce over-penalization of logic-preserving edits, code-aware similarity is measured using CodeBLEU~\cite{ren2020codebleu}. In the finalized implementation, the data-flow component is excluded for stability, and CodeBLEU is computed using the n-gram, weighted n-gram, and syntax components only. The three are weighted equally, \((1/3,1/3,1/3,0)\), and C submissions are evaluated under the C++ CodeBLEU configuration for toolchain stability.
Consecutive-turn churn is
\vspace{-0.3em}
\begin{equation}
\mathrm{Churn}^{\mathrm{CB}}_{u,p,k}
=
1-\mathrm{CodeBLEU}(c_{u,p,k-1},c_{u,p,k}),
\qquad k\ge 2,
\end{equation}
and, for solved trajectories, convergence toward the first accepted solution \(c^{\mathrm{AC}}_{u,p}\) is measured for every attempt other than that submission itself as
\vspace{-0.3em}
\begin{equation}
\mathrm{Conv}^{\mathrm{CB}}_{u,p,k}
=
\mathrm{CodeBLEU}(c^{\mathrm{AC}}_{u,p},c_{u,p,k}).
\end{equation}

\subsection{Inference Configuration}
\label{sec:inference_config}
\tabboxref{inference_settings} summarizes the inference configuration. Gemini 2.5 Pro cannot be invoked without internal reasoning, and an unbounded reasoning allowance causes it to exhaust its output budget before emitting a response on a substantial share of prompts. An explicit reasoning budget is therefore set for every Gemini request; a paired check on ten attempts (nine comparable pairs), scored with and without that budget, confirmed it does not alter the judgement. All other judges use their provider defaults. Before the evaluation run, each provider was asked which model it actually serves under the requested identifier. Three identifiers were version-pinned or resolved unchanged; the DeepSeek identifier \texttt{deepseek-reasoner} resolved to \texttt{deepseek-v4-flash}, so the run pins that explicit identifier, and the identifier each provider reports is recorded with every response.

\begin{table}[h]
\caption{Inference configuration for the API-based \emph{LLM-as-a-Judge} pipeline}
\setlength{\arrayrulewidth}{0.2mm}
\setlength{\tabcolsep}{2pt}
\renewcommand{\arraystretch}{1}
\label{inference_settings}
\begin{tabular}{l||l}
\hline \hline
\textbf{Settings}                  & \textbf{Value/Description}        \\ \hline \hline
Judge models & \shortstack {\texttt{gpt-5.2-2025-12-11}, \texttt{gemini-2.5-pro}, \\ \texttt{claude-sonnet-4-5-20250929}, \\ \texttt{deepseek-v4-flash}} \\ \hline 
Input conditions & \textsc{blind}, \textsc{prefix12k}, \textsc{nonblind} (Section~\ref{sec:approach_data}) \\ \hline
Context type & \shortstack{Per condition: none, the first 12,000 characters of \\the aggregated logs, or the anchored context \\(Section~\ref{sec:approach_data})} \\ \hline
Output format & Strict JSON schema \\ \hline
Output fields & \(p_{\mathrm{ac}}, s_{\mathrm{algo}}, s_{\mathrm{robust}}, r\) \\ \hline
Score scale & \(s_{\mathrm{algo}}, s_{\mathrm{robust}} \in \{1,\dots,5\}\) \\ \hline
Temperature & 0 (all judges) \\ \hline
Max output tokens & \shortstack {OpenAI/DeepSeek: provider default;\; Gemini: \\2048 (main), 4096 (repair);\; Claude: 1024} \\ \hline 
\shortstack{Gemini reasoning \\budget} & \shortstack{1,024 thinking tokens (a soft target, not a hard \\cap; observed up to 1,395 in practice); confirmed \\not to alter the judgement (mean absolute \\difference 0.036, 8 of 9 pairs within 0.10)} \\ \hline
Code input budget & \texttt{MAX\_CODE\_CHARS}=12000 \\ \hline 
Prompt/context budget & \texttt{MAX\_PROMPT\_CHARS}=12000 \\ \hline 
\shortstack{Request pacing \\ / checkpointing} & \shortstack{\texttt{SLEEP\_SECONDS}=0.2, \texttt{SAVE\_EVERY}=10; \\resume-safe checkpointing enabled} \\ \hline
Retry policy & \shortstack {Bounded retry with backoff; retry-only-failed \\with UPSERT by \texttt{attempt\_id}} \\ \hline
Evaluation folds & \shortstack{Grouped 10-fold CV by \((u,p)\) (Section~\ref{sec:approach_analysis})} \\ \hline 
Threshold criterion & \shortstack{MCC selected on the other 9 folds, \\applied unchanged to the held-out fold} \\ \hline
\shortstack{Per-request latency \\(aggregate)} & \shortstack{About 1.5s (fastest judge) to about 90s \\(slowest), one to two orders of magnitude, \\over the same 517 attempts} \\ \hline
\shortstack{OpenAI latency \\(s/request)} & \textsc{blind} 1.5, \textsc{prefix12k} 1.8, \textsc{nonblind} 1.8 \\ \hline
\shortstack{Claude latency \\(s/request)} & \shortstack{\textsc{blind} 2.7, \textsc{nonblind} 2.9; \textsc{prefix12k} not \\cleanly measurable (pass split by provider \\credit exhaustion)} \\ \hline
\shortstack{Gemini latency \\(s/request)} & \shortstack{\textsc{blind} 10.3, \textsc{prefix12k} 10.7; \textsc{nonblind} \\not cleanly measurable (pass interrupted by \\provider quota)} \\ \hline
\shortstack{DeepSeek latency \\(s/request)} & \textsc{blind} 77.6, \textsc{nonblind} 83.5, \textsc{prefix12k} 92.2 \\ \hline
ECE bins & 15 \\ \hline \hline
\end{tabular}
\end{table}

\subsection{Perturbation Study Design}
\label{sec:perturbation_design}
The evaluation above scores judges against the contest verdict on unmodified attempts. This section adds a second, independent check: whether a judge's score responds to a defect that the corpus does not already contain, against a matched edit that provably changes nothing.

\textbf{Seeds.} The 134 accepted programs written in C++, C, or Java (57, 39, and 38) serve as mutation seeds, since all four operators below apply in these languages. The 24 accepted Python and Go programs are not used; in Python, the narrowed-type defect cannot occur.

\textbf{Semantic operators.} Four operators each introduce a single defect of a kind the contest judge system is already known to reject, ordered here by how difficult the defect is to notice, subtlest first:
\begin{itemize}
\item \textbf{TYPE\_NARROW} (severity 1): a \texttt{long long} declaration is narrowed to \texttt{int}. The program still compiles and remains correct below 32 bits, failing only once a value exceeds that range, which is exactly what contest input bounds are designed to reach.
\item \textbf{ROR\_BOUNDARY} (severity 2): a relational operator at a boundary check is tightened or loosened, \texttt{<=} to \texttt{<} or the reverse, an off-by-one at that boundary.
\item \textbf{LOOP\_SHORTEN} (severity 3): a loop bound \texttt{i < n} becomes \texttt{i < n - 1}, dropping the last iteration.
\item \textbf{ARITH\_FLIP} (severity 4): an addition becomes a subtraction, \texttt{a + b} to \texttt{a - b}, a change wrong on essentially every input.
\end{itemize}
Each operator changes exactly one occurrence, the first one it finds outside comments, preprocessor directives, and string or character literals, so every mutation is auditable by eye against the line it replaces.

\textbf{Matched control arm.} Two operators edit the same kind of construct without changing what the program computes, so that a judge's response to a real defect can be told apart from a response to the fact that code changed at all: a post-increment is rewritten to a pre-increment inside a for-loop header (\texttt{i++} to \texttt{++i}, behavior-identical in C, C++, and Java, since the result of that expression is discarded there), and a comment line is inserted at a fixed point.

\textbf{Yield.} The operators reach different shares of the 134 seeds, giving 97 TYPE\_NARROW, 91 ROR\_BOUNDARY, 71 LOOP\_SHORTEN, and 89 ARITH\_FLIP mutants (348 semantic) and 73 post-increment and 134 comment controls (207); all 555 passed verification. Guard-context exposure is measured: 35 of the 348 semantic mutants sit in a guard condition where the edit may not change behavior, for instance a tightened boundary check that excludes a value the problem's constraints already rule out.

\textbf{Judge requests and the primary subset.} All 555 mutants are scored by all four judges under the \textsc{blind} condition, since a mutant is a synthetic program with no participant history of its own to attach; 2,220 requests in all. The primary analysis is restricted to the subset in which the same judge scored the unmutated seed at or above 0.5 under the same condition, since a judge that misjudged the original program can be neither credited nor blamed for what it does to the mutant; that subset holds 1,507 of the 2,220 judge-mutant pairs.

\section{Experimental Results}
\label{Results}
Judges are named by provider;~\tabboxref{inference_settings} gives the model identifiers. Results are reported over all 517 attempts under grouped 10-fold CV (Section~\ref{sec:approach_analysis}), for each of the three input conditions defined in Section~\ref{sec:approach_data}. All intervals are 95\% trajectory-clustered bootstrap intervals from 2,000 resamples; the tables give one for every value, and the text gives one wherever a comparison rests on it. Two judges are described as different only where their intervals do not overlap, and two conditions only where the interval on their paired difference excludes zero.

\subsection{LLM-as-a-Judge Evaluation Results}
\label{results_judge}

\subsubsection{Effect of Input Specification}
\label{sec:statement_effect}
The earlier canonical request carried a problem identifier in place of the problem statement, so a judge assessing algorithmic adequacy had not seen the specification it was judging against. Supplying the full statement rebuilds the ranking: over the same 517 attempts at the same temperature, the rank correlation between the two orderings is 0.3021 [0.168, 0.424], and the mean judged probability moves from 0.1840 [0.158, 0.214], well below the base rate of 0.3056, to 0.3561 [0.311, 0.409], slightly above it. Under \textsc{nonblind}, out-of-fold ROC-AUC rises for every judge, from 0.130 [0.061, 0.202] for Gemini to 0.295 [0.214, 0.381] for OpenAI, every gain's interval excluding zero (~\tabboxref{tab:statement_effect}). The \textsc{blind} condition removes participant context entirely, so a gain there cannot come from better context: three judges gain under it as well, from 0.095 [0.022, 0.172] for Gemini to 0.249 [0.177, 0.319] for DeepSeek, while Claude's gain, 0.065 [$-0.023$, 0.158], separates from zero only once participant context is added (0.141 [0.058, 0.227] under \textsc{nonblind}). For OpenAI, Claude, and Gemini, model, corpus, labels, and temperature are identical across the two runs. DeepSeek's earlier run requested an alias whose served model cannot be recovered (Section~\ref{sec:inference_config}), so its gain may include a change of model; the gains of the other three judges carry the comparison on their own, and without DeepSeek the ordering is rebuilt at least as fully, with a rank correlation of 0.2617 [0.123, 0.391] across those three judges.

With the problem statement held fixed, adding participant context improves discrimination for two of the four judges. Moving from \textsc{blind} to \textsc{nonblind} raises out-of-fold ROC-AUC for Claude by 0.0762 [0.034, 0.120] and for OpenAI by 0.0672 [0.030, 0.106], and by 0.0903 [0.039, 0.143] and 0.0805 [0.040, 0.128] on the 471 attempts whose context is anchored or segmented (Section~\ref{sec:approach_data}); for DeepSeek and Gemini the change is not detectable either way, its interval including zero. The benefit of context is therefore judge-specific and is reported per judge, with \textsc{nonblind} as the main reporting condition because it gives every judge the most relevant evidence.

\begin{table*}[t]
\centering
\caption{ROC-AUC by input condition, out-of-fold, all 517 attempts, with 95\% intervals}
\setlength{\arrayrulewidth}{0.2mm}
\setlength{\tabcolsep}{37pt}
\renewcommand{\arraystretch}{1}
\label{tab:statement_effect}
\small
\begin{tabular}{@{}l||c||c||c@{}}
\hline
Judge & \textsc{blind} & \textsc{prefix12k} & \textsc{nonblind} \\
\hline \hline
Claude & 0.6224 [0.557, 0.689] & 0.6815 [0.612, 0.752] & 0.6985 [0.633, 0.763] \\ \hline
DeepSeek & 0.9020 [0.863, 0.936] & 0.9357 [0.904, 0.961] & 0.9236 [0.890, 0.954] \\ \hline
Gemini & 0.7345 [0.675, 0.791] & 0.7574 [0.704, 0.811] & 0.7691 [0.719, 0.817] \\ \hline
OpenAI & 0.7535 [0.693, 0.815] & 0.7975 [0.739, 0.853] & 0.8208 [0.767, 0.874] \\ \hline
\hline
\end{tabular}
\end{table*}

Against \textsc{prefix12k}, which reproduces the context rule of the earlier run and fills its 12,000-character cap on every attempt, \textsc{nonblind} sends 6,821 characters on average, about 43\% less, and no judge separates the two on all 517 attempts; on the 471 evidence-backed attempts, only OpenAI gains, by 0.0516 [0.018, 0.088]. Judge performance is reported conditionally on the input specification throughout this section, since the specification determines what a score measures.

\subsubsection{Discrimination (ROC-AUC, PR-AUC)}
\begin{figure}[h]
\centering
\subfloat[ROC-AUC]{\includegraphics[width=1.74in]{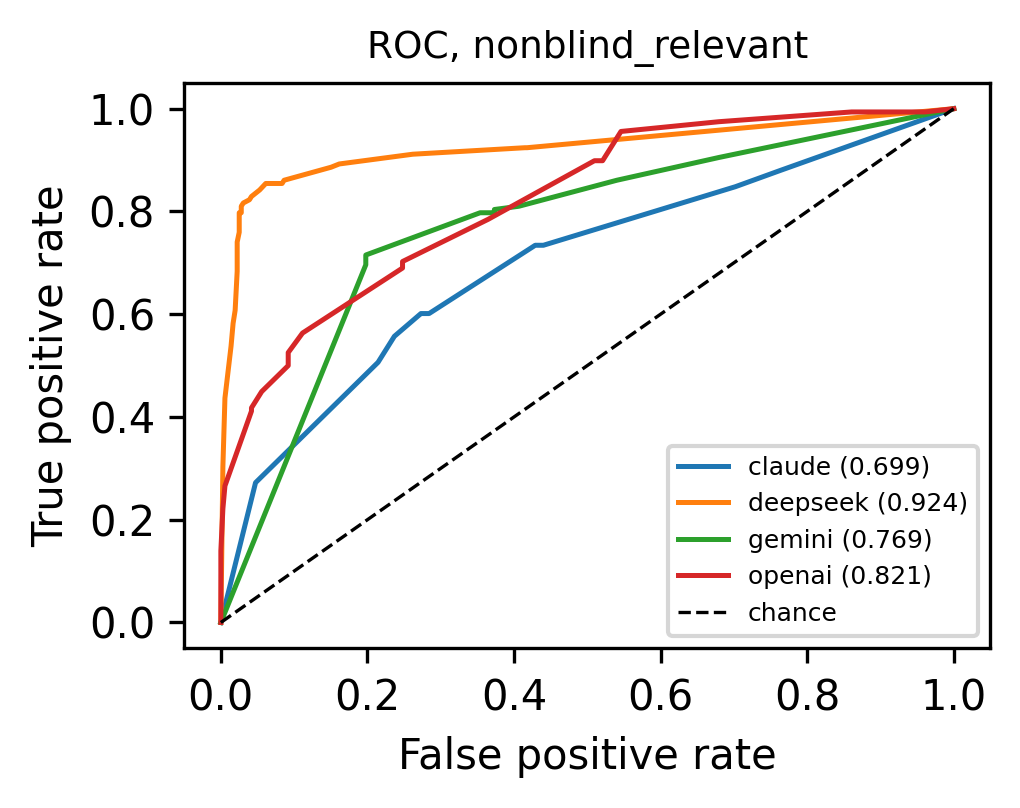}%
\label{fig_roc}}
\hfil
\subfloat[PR-AUC]{\includegraphics[width=1.74in]{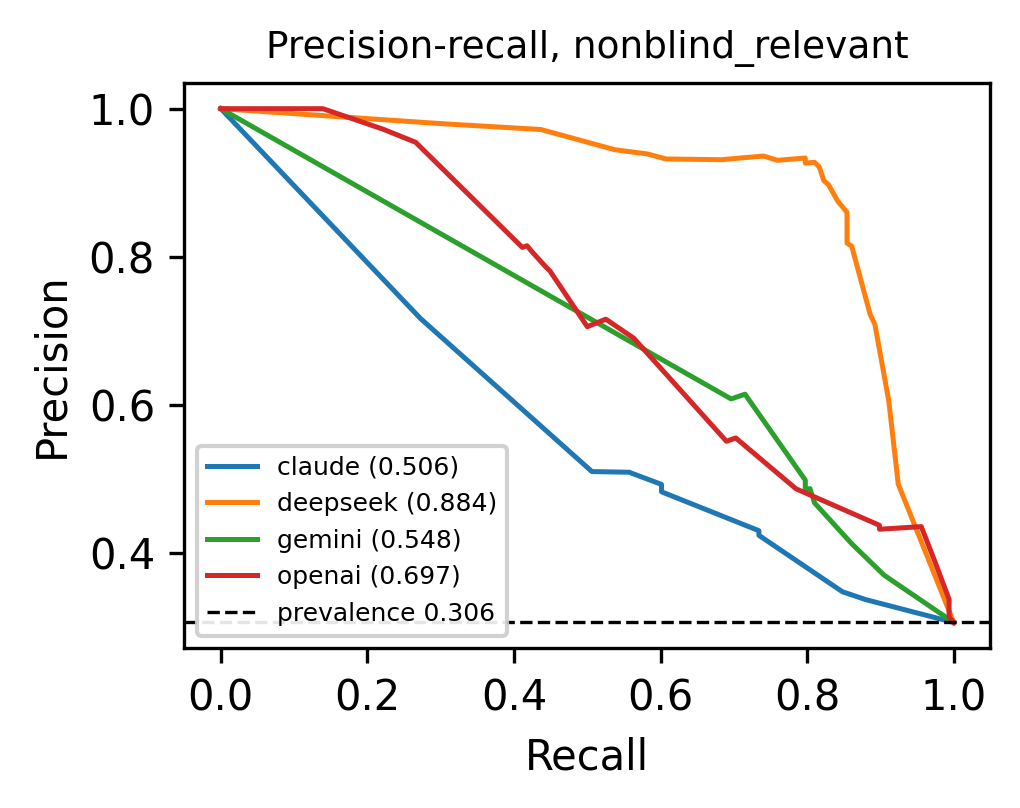}%
\label{fig_pr}}
\caption{(a) ROC and (b) PR curves per judge, out-of-fold, \textsc{nonblind}, 517 attempts}
\label{auc_curves}
\end{figure}

DeepSeek ranks best (\tabboxref{judge_all_metrics}), with the highest ROC-AUC (0.9236) and PR-AUC (0.8844), its intervals clear of the other three judges on both. OpenAI (ROC-AUC 0.8208; PR-AUC 0.6969), Gemini (0.7691; 0.5480) and Claude (0.6985; 0.5060) follow in that order on both measures, but adjacent intervals overlap, so among them only OpenAI's lead over Claude on ROC-AUC separates. All four ROC-AUC intervals exclude chance. The curves in~\figboxref{auc_curves} show where these differences arise. DeepSeek's ROC curve~\figboxref{fig_roc} reaches a TP rate near 0.8 at a FP rate below 0.05, and its PR curve~\figboxref{fig_pr} holds precision above 0.9 up to a recall near 0.8. The OpenAI and Gemini curves cross more than once in both panels, so which of the two is better depends on the operating point, in line with their overlapping intervals. Gemini's ROC curve opens with a single straight segment to about (0.2, 0.7), its 181 predictions tied at exactly 1.00, which no threshold can split: the scale-validity effect of Section~\ref{sec:scale_validity}, seen directly. Every curve stays above chance, and every PR curve above the prevalence of 0.306 until full recall. Averaged, the two rubric scores rank attempts better than the probability for DeepSeek (ROC-AUC difference 0.0424 [0.016, 0.071]) and worse for OpenAI ($-0.0295$ [$-0.057$, $-0.002$]), with no detectable difference for Claude and Gemini.

\begin{table*}[t]
\centering
\caption{\emph{LLM-as-a-Judge} performance, \textsc{nonblind} condition, out-of-fold over all 517 attempts, with 95\% trajectory-clustered bootstrap intervals}
\setlength{\arrayrulewidth}{0.2mm}
\setlength{\tabcolsep}{10.5pt}
\renewcommand{\arraystretch}{1}
\label{judge_all_metrics}
\begin{tabular}{l||c||c||c||c||c||c||c}
\hline \hline
\textbf{Judge} & \textbf{ROC-AUC $\uparrow$} & \textbf{PR-AUC $\uparrow$} & \textbf{$\mathcal{L}$ $\downarrow$} & \textbf{$B$ $\downarrow$} & \textbf{ECE$_{15}$ $\downarrow$} & \textbf{MCC $\uparrow$} & \textbf{MCC$_{\mathrm{oof}}$ $\uparrow$} \\ \hline \hline
\texttt{OpenAI} & \shortstack{0.8208\\{\scriptsize[0.767, 0.874]}} & \shortstack{0.6969\\{\scriptsize[0.600, 0.798]}} & \shortstack{0.5477\\{\scriptsize[0.438, 0.667]}} & \shortstack{0.1672\\{\scriptsize[0.134, 0.205]}} & \shortstack{0.1261\\{\scriptsize[0.089, 0.175]}} & \shortstack{0.4810\\{\scriptsize[0.414, 0.595]}} & \shortstack{0.4396\\{\scriptsize[0.338, 0.543]}} \\ \hline
\texttt{DeepSeek} & \shortstack{0.9236\\{\scriptsize[0.890, 0.954]}} & \shortstack{0.8844\\{\scriptsize[0.828, 0.933]}} & \shortstack{0.5236\\{\scriptsize[0.323, 0.762]}} & \shortstack{0.0717\\{\scriptsize[0.049, 0.101]}} & \shortstack{0.0627\\{\scriptsize[0.043, 0.097]}} & \shortstack{0.8147\\{\scriptsize[0.761, 0.874]}} & \shortstack{0.7822\\{\scriptsize[0.701, 0.850]}} \\ \hline
\texttt{Gemini} & \shortstack{0.7691\\{\scriptsize[0.719, 0.817]}} & \shortstack{0.5480\\{\scriptsize[0.448, 0.665]}} & \shortstack{2.7769\\{\scriptsize[2.216, 3.340]}} & \shortstack{0.3045\\{\scriptsize[0.251, 0.355]}} & \shortstack{0.3036\\{\scriptsize[0.250, 0.358]}} & \shortstack{0.4978\\{\scriptsize[0.404, 0.592]}} & \shortstack{0.4978\\{\scriptsize[0.404, 0.592]}} \\ \hline
\texttt{Claude} & \shortstack{0.6985\\{\scriptsize[0.633, 0.763]}} & \shortstack{0.5060\\{\scriptsize[0.395, 0.641]}} & \shortstack{1.6549\\{\scriptsize[1.229, 2.128]}} & \shortstack{0.2807\\{\scriptsize[0.219, 0.342]}} & \shortstack{0.2647\\{\scriptsize[0.198, 0.332]}} & \shortstack{0.3233\\{\scriptsize[0.244, 0.453]}} & \shortstack{0.2120\\{\scriptsize[0.087, 0.356]}} \\ \hline \hline
\end{tabular}
\end{table*}

\subsubsection{Probabilistic scoring ($\mathcal{L}$, $B$)}
On Brier, DeepSeek is lowest ($B=0.0717$, against 0.1672 for OpenAI, intervals separating); on $\mathcal{L}$, DeepSeek and OpenAI are not distinguishable (0.5236 and 0.5477). Gemini and Claude have much larger $\mathcal{L}$ (2.7769 and 1.6549), consistent with poorly calibrated or overconfident wrong predictions. Calibration error across judges and conditions ranges from 0.052 to 0.374, so this is not confined to one judge or setting.

\subsubsection{Calibration (ECE)}
\begin{figure}[h]
\centering
\includegraphics[width=3in]{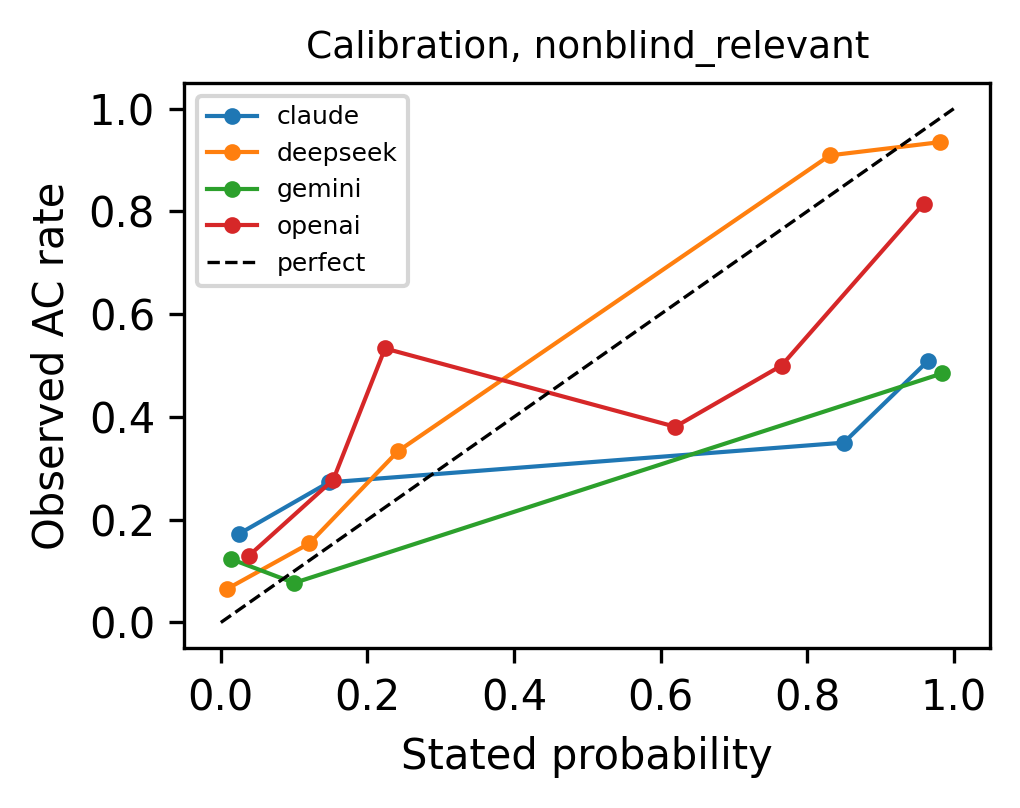}
\caption{Judge-level reliability: observed acceptance rate against mean predicted probability per bin, out-of-fold, \textsc{nonblind} condition}
\label{fig_calibration}
\end{figure}

ECE$_{15}$ separates the judges into two tiers: DeepSeek (0.0627) and OpenAI (0.1261), whose intervals overlap, are better calibrated than Claude (0.2647) and Gemini (0.3036), whose intervals overlap with each other but with neither of the first two. Calibration matters wherever a judge's probability is read at face value, for instance when a threshold set on one sample is applied to another: lower ECE indicates that predicted probabilities align more closely with empirical acceptance frequencies on held-out attempts. The co-creation signal of Section~\ref{co-creation} uses each judge's ranks only, so no order-preserving recalibration changes it. The reliability diagram in~\figboxref{fig_calibration} shows the same split: DeepSeek's curve tracks the diagonal closely throughout, while Claude and Gemini both compress toward the middle at the high end, so a stated probability above 0.85 yields an observed acceptance rate of only 0.4 to 0.5 for either judge.

\subsubsection{Thresholded decision quality (MCC, out-of-fold)}
MCC$_{\mathrm{oof}}$ pools the out-of-fold decisions of all ten folds (Section~\ref{sec:approach_analysis}). DeepSeek achieves the highest MCC$_{\mathrm{oof}}$ (0.7822), with an interval clear of the other three; Gemini (0.4978, identical to its in-fold value) and OpenAI (0.4396) overlap, and Claude (0.2120) separates from Gemini but not from OpenAI. The gap between MCC and MCC$_{\mathrm{oof}}$ in~\tabboxref{judge_all_metrics} shows how much of a judge's apparent decision quality depended on the fold it was measured on; that gap is examined directly in Section~\ref{sec:scale_validity}.

\subsubsection{Cross-metric comparison}
Two cross-metric comparisons, illustrated in~\figboxref{cross_metric_comparison}, test whether one favorable measurement transfers to another. Panel (a) compares discrimination (ROC-AUC/PR-AUC) against out-of-fold decision quality (MCC$_{\mathrm{oof}}$): favorable discrimination transfers to decision quality, with DeepSeek leading on both and its intervals clear of the others, while OpenAI and Gemini change places between the two measures only within overlapping intervals. Panel (b) compares calibration (ECE$_{15}$) against probabilistic scoring ($\mathcal{L}$/$B$). Here the point-estimate order is the same on all three measures, DeepSeek, OpenAI, Claude, Gemini, and on each the intervals separate DeepSeek and OpenAI from Claude and Gemini, so probability quality transfers across measures at the level the intervals support.

\begin{figure}[h]
\centering
\subfloat[Ranking vs.\ decision\label{discrimination_vs_mcc}]{
    \begin{minipage}[t]{0.47\linewidth}
        \centering
        \input{Graphs_Figures/comp1}\vspace{-12pt}
    \end{minipage}
}
\hfill
\subfloat[Calibration vs.\ scoring\label{calib_vs_probabilistic}]{
    \begin{minipage}[t]{0.47\linewidth}
        \centering
        \input{Graphs_Figures/comp2}\vspace{-6pt}
    \end{minipage}
}
\caption{Cross-metric comparison, \textsc{nonblind}: (a) ROC-AUC and PR-AUC against MCC$_{\mathrm{oof}}$; (b) ECE$_{15}$, $\mathcal{L}$, and $B$, min-max normalized so higher is better (text uses raw values)}
\label{cross_metric_comparison}
\end{figure}

\subsubsection{Scale Validity}
\label{sec:scale_validity}
A number in [0,1] can satisfy the schema without saying much. Across the three input conditions, the four judges emit as few as 8 and as many as 31 distinct probability values, collapsing a grid of 99 candidate decision thresholds to as few as 6 effective rules. Under \textsc{nonblind}, Claude, DeepSeek, Gemini, and OpenAI emit 11, 31, 10, and 22 distinct values, leaving 10, 28, 8, and 20 effective rules of 99. Concentration compounds the effect: Gemini places 181 of its 517 predictions (35.0\%) at exactly 1.00 and DeepSeek 221 (42.7\%) at exactly 0.00, in a corpus where 69.4\% of attempts are not accepted. Coarse scales also make decision quality depend on where the threshold is selected: under \textsc{blind}, moving from in-fold to out-of-fold thresholds lowers MCC by 0\% for Gemini (0.4061 both), 1.1\% for DeepSeek, 45.7\% for Claude, and 62.5\% for OpenAI, the one judge whose two intervals do not overlap. Schema compliance and scale validity are therefore separate properties: every judge returns a well-formed probability, but not an equally informative one.

\subsubsection{Metric-Ranking Invariance}
\label{sec:metric_invariance}
Which judge ranks best depends on which metric and which threshold-selection method is used, a practical consequence of scale validity, not a separate finding. Under the in-fold threshold, OpenAI's point estimate sits above Claude's in the \textsc{blind} condition (MCC 0.3741 [0.335, 0.486] against 0.2184 [0.134, 0.357]); under out-of-fold selection on the 369 anchored attempts, in the same condition, it falls below (MCC$_{\mathrm{oof}}$ 0.0729 [$-0.091$, 0.245] against 0.1463 [0.007, 0.283]). Neither difference separates, which is the point: a single ranking would present an unstable order as a fixed one. The perturbation study, reported in Section~\ref{sec:perturbation_results}, separates the judges more clearly: counting a score drop of more than 0.05 as a reaction, DeepSeek, the most accurate judge on discrimination, reacted to 27 of 180 behavior-preserving edits, and Claude, the least accurate, to 1 of 129. No judge is strongest on every measure, so two criteria are used together, not one: a discrimination interval that excludes chance, and an arm-discrimination interval that excludes chance on the perturbation check. Calibration follows the same principle: with no established cut-off for this setting, its value and consequence for threshold transfer are reported directly.

\subsubsection{Pairwise and multi-rater agreement (Cohen's $\kappa$, Fleiss' $\kappa$)}
Inter-judge reliability is evaluated over all 517 attempts under each of the three input conditions, with each judge's continuous output cut at its own median, a rank-based cut that places judges of different output resolution on equal terms; since the conditions differ only in the participant context supplied, any resulting difference is attributable to that context alone (~\tabboxref{kappa_summary}).

\begin{table*}[t]
\centering
\caption{Inter-judge agreement across input conditions, each judge cut at its own median, with 95\% trajectory-clustered bootstrap intervals. Max and Min repeat Panel B values, whose intervals are given there. Judge codes: C=Claude, D=DeepSeek, G=Gemini, O=OpenAI}
\label{kappa_summary}
\setlength{\arrayrulewidth}{0.2mm}
\setlength{\tabcolsep}{3.5pt}
\renewcommand{\arraystretch}{1}
\begin{tabular}{l|c|c|c|c|c|c||l|c|c|c}
\hline
\multicolumn{7}{c||}{\textbf{Panel A: condition-level summary}} & \multicolumn{4}{c}{\textbf{Panel B: pairwise Cohen's $\kappa$ by condition}} \\
\hline \hline
Condition & Fleiss' $\kappa$ & Mean $\kappa$ & Mean $\rho$ & $\rho-\kappa$ & Max $\kappa$ (pair) & Min $\kappa$ (pair) & Pair & \textsc{blind} & \textsc{prefix12k} & \textsc{nonblind} \\
\hline
\textsc{blind} & \shortstack{0.3714\\{\scriptsize[0.299, 0.441]}} & \shortstack{0.3718\\{\scriptsize[0.300, 0.441]}} & \shortstack{0.4863\\{\scriptsize[0.410, 0.553]}} & \shortstack{0.1145\\{\scriptsize[0.078, 0.151]}} & 0.4844 (G-O) & 0.2590 (C-D) & C-D & \shortstack{0.2590\\{\scriptsize[0.164, 0.354]}} & \shortstack{0.2497\\ {\scriptsize[0.148, 0.368]}} & \shortstack{0.3494\\{\scriptsize[0.246, 0.461]}} \\ \hline
\textsc{prefix12k} & \shortstack{0.4178\\{\scriptsize[0.343, 0.493]}} & \shortstack{0.4199\\{\scriptsize[0.347, 0.495]}} & \shortstack{0.5414\\{\scriptsize[0.474, 0.606]}} & \shortstack{0.1216\\{\scriptsize[0.082, 0.156]}} & 0.5021 (C-O) & 0.2497 (C-D) & C-G & \shortstack{0.4250\\{\scriptsize[0.312, 0.526]}} & \shortstack{0.4224\\{\scriptsize[0.327, 0.526]}} & \shortstack{0.4246\\{\scriptsize[0.319, 0.534]}} \\ \hline
\textsc{nonblind} & \shortstack{0.4353\\{\scriptsize[0.366, 0.510]}} & \shortstack{0.4366\\{\scriptsize[0.368, 0.512]}} & \shortstack{0.5533\\{\scriptsize[0.488, 0.614]}} & \shortstack{0.1167\\{\scriptsize[0.075, 0.155]}} & 0.5589 (G-O) & 0.3494 (C-D) & C-O & \shortstack{0.3232\\{\scriptsize[0.221, 0.414]}} & \shortstack{0.5021\\{\scriptsize[0.407, 0.588]}} & \shortstack{0.4567\\{\scriptsize[0.354, 0.544]}} \\ \hline
\multicolumn{7}{c||}{} & D-G & \shortstack{0.3771\\{\scriptsize[0.270, 0.477]}} & \shortstack{0.4269\\{\scriptsize[0.331, 0.526]}} & \shortstack{0.4149\\{\scriptsize[0.321, 0.510]}} \\ \cline{8-11}
\multicolumn{7}{c||}{}  & D-O & \shortstack{0.3621\\{\scriptsize[0.260, 0.459]}} & \shortstack{0.4255\\{\scriptsize[0.309, 0.542]}} & \shortstack{0.4152\\{\scriptsize[0.306, 0.532]}} \\ \cline{8-11}
\multicolumn{7}{c||}{}  & G-O & \shortstack{0.4844\\{\scriptsize[0.378, 0.579]}} & \shortstack{0.4927\\{\scriptsize[0.392, 0.594]}} & \shortstack{0.5589\\{\scriptsize[0.468, 0.643]}} \\ \hline
\hline
\end{tabular}
\end{table*}

\begin{figure}[h]
\centering
\includegraphics[width=3in]{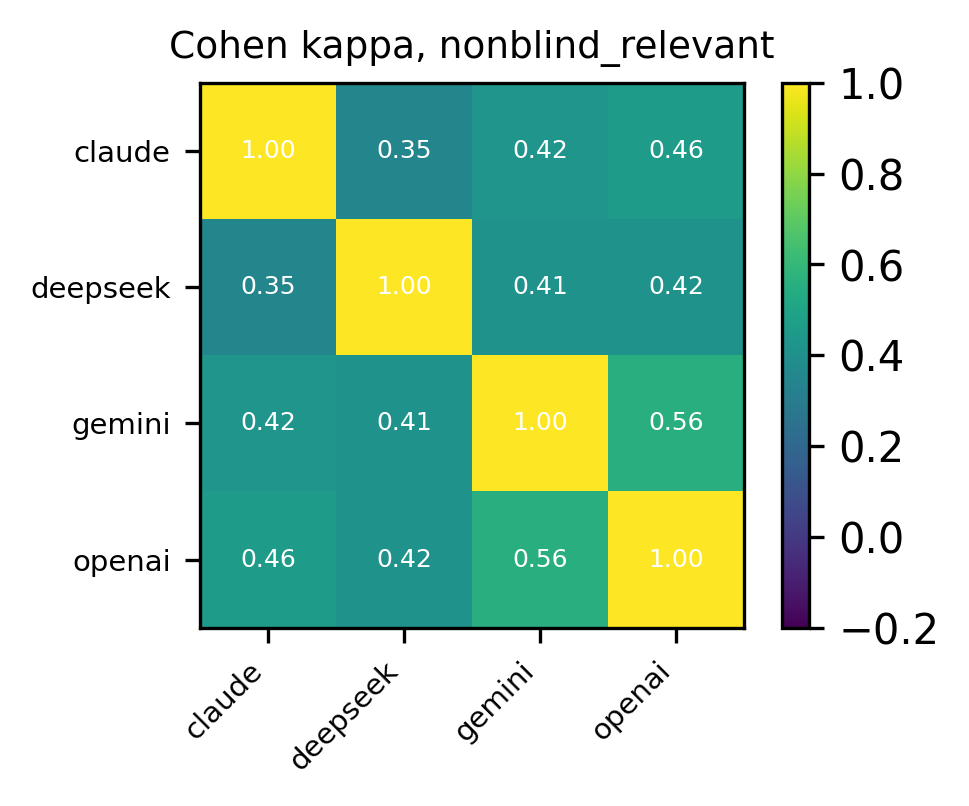}
\caption{Pairwise Cohen's $\kappa$, \textsc{nonblind} condition, all 517 attempts, each judge cut at its own median}
\label{kappa_heatmap}
\end{figure}

Agreement is higher with relevance-selected context than with none: on the same resampled attempts, Fleiss' $\kappa$ rises by 0.0639 [0.0005, 0.1336] from \textsc{blind} to \textsc{nonblind}, and mean rank agreement by 0.0670 [0.0158, 0.1233], while \textsc{prefix12k} lies between the two without separating from either on $\kappa$. No pairwise agreement is negative, and no pair's interval reaches zero, under any condition (Panel B). Claude-DeepSeek is the lowest pair under every condition, and the leading pairs' intervals overlap in every condition. Mean $\rho$ exceeds mean $\kappa$ by 0.11 to 0.12 in every condition, no more than reducing each ranking to two halves produces on its own, so the residual disagreement concerns which attempts are better, not merely where each judge cuts. \figboxref{kappa_heatmap} shows the \textsc{nonblind} pairs; Panel B gives all three conditions.         

One further result bears on how much this residual disagreement should be trusted: agreement across instruments is only half of consistency. Under repeated application, 17 consecutive-attempt pairs in the corpus consist of character-identical resubmissions, and the judged probability moved between them at temperature zero, evidence that some of it reflects the judges themselves and not only the input they are given. The four judges are therefore not interchangeable, and their divergence is reported as a limit with guidance attached: DeepSeek suits ranking and triage, and Claude suits uses where the score must not move on edits that leave behavior unchanged, on the evidence given in Section~\ref{sec:metric_invariance}.

\subsubsection{Integrated performance perspective}
\label{integrated}
Pooled metrics can also hide how a judge fails. DeepSeek's own errors, examined by type, are mostly confident FN: of its 47 out-of-fold errors under \textsc{nonblind}, 30 reject attempts the contest accepted, 27 of them at an acceptance probability of 0.2 or below, a different risk from a judge that is merely uncertain. The per-problem breakdown makes such risks visible where pooled figures hide them: on one problem, Gemini's ROC-AUC falls to 0.0614 under \textsc{nonblind}, an ordering inverted within that problem although its pooled discrimination holds. Combining the four judges tests this directly: an unweighted percentile-rank panel scores 0.8734 [0.831, 0.913] ROC-AUC under \textsc{nonblind}, 0.070 [0.059, 0.082] above the mean member (0.8030 [0.767, 0.841]) but below the single strongest member, DeepSeek (0.9236; difference $-0.0502$ [$-0.0883$, $-0.0123$], excluding zero). The panel offers a safe default when the best judge for a task is not yet known; knowing that judge offers more still. Overall, no single judge leads on every measure, so a multi-criteria view is what matches each judge to the use it has been validated for.

\subsubsection{Decision performance vs.\ agreement-based reliability}
\figboxref{judge_agreement_overlay} relates each judge's MCC$_{\mathrm{oof}}$ to its mean pairwise Cohen's $\kappa$ to check whether stronger decisions come with broader agreement. They do not: DeepSeek pairs the strongest decision quality with the lowest mean pairwise $\kappa$ (0.3932 [0.303, 0.487], against 0.4102 to 0.4769 for the others), with all four intervals overlapping.

\begin{figure}[h]
\centering
\makebox[\linewidth][c]{%
\subfloat[Raw decision and agreement\label{mcc_vs_kappa}]{%
    \begin{minipage}[t]{0.485\linewidth}
        \centering
        \resizebox{\linewidth}{!}{\input{Graphs_Figures/comp3}}
    \end{minipage}%
}%
\hspace{0.005\linewidth}%
\subfloat[Relative judge profiles\label{mcc_kappa_profile}]{%
    \begin{minipage}[t]{0.485\linewidth}
        \centering
        \resizebox{\linewidth}{!}{\input{Graphs_Figures/comp4}}
    \end{minipage}%
}%
}
\caption{Decision quality and agreement per judge, \textsc{nonblind}: (a) MCC$_{\mathrm{oof}}$ and mean pairwise $\kappa$; (b) the same, min-max normalized (text uses raw values)}
\label{judge_agreement_overlay}
\end{figure}

\subsubsection{Robustness Checks}
\label{sec:sensitivity}
The main findings hold under three independent checks. First, clustering: resampling the 13 problems in place of the trajectories, a stricter assumption that treats all attempts on a problem as dependent, widens the intervals by 1.76$\times$ on average, yet every judge still exceeds chance, Claude most narrowly at $[0.5887, 0.8126]$; all eight condition-level findings keep their status; and DeepSeek's lead over Gemini and Claude still separates (DeepSeek [0.8687, 0.9683], Gemini [0.6811, 0.8507]). Only the two closest comparisons, DeepSeek over OpenAI ([0.7205, 0.9147]) and OpenAI over Claude, become ties. Second, problem weighting: giving every problem equal weight under \textsc{nonblind} changes each judge's ROC-AUC by less than 0.05 (macro-averaged 0.9161 against 0.9236 for DeepSeek), so the pooled figures are not driven by the largest problems. Third, prompt-design contamination: under the same condition, the 38 trajectories (116 attempts) never inspected while the judge prompt was designed score at least as high as the full corpus for every judge, by 0.0005 to 0.0556, with every interval excluding 0.5 and a similar prevalence (0.2931 against 0.3056).

\subsection{Perturbation Study Results}
\label{sec:perturbation_results}
Results are reported on the primary subset defined in Section~\ref{sec:perturbation_design}, with 95\% intervals from 2,000 bootstrap resamples of seeds, since the mutants of one program are not independent observations. For each mutant, the shift $\Delta$ is its judged probability minus that of its seed, both under \textsc{blind}. Three properties of the score are measured: arm discrimination $A_j$, the probability that a control mutant's shift exceeds a semantic mutant's, which is the ROC-AUC of the shift as a separator of the two arms; detection specificity, how often each arm reacts, counting a drop of more than 0.05 as a reaction; and the severity gradient $\rho_j$, the Spearman correlation between operator severity and the size of the drop. Every shift is measured from a seed the judge itself scored as acceptable: the primary subset keeps 84 to 118 seeds for DeepSeek, Gemini and Claude, and 59 for OpenAI, which scored 75 of the 134 accepted seeds below 0.5.

\begin{table}[h]
\centering
\caption{Perturbation results, primary subset, \textsc{blind}, 95\% seed-clustered intervals. Reactions: mutants whose judged probability drops by more than 0.05, of those scored; semantic arm (sem.) above control arm (ctl.)}
\setlength{\arrayrulewidth}{0.2mm}
\setlength{\tabcolsep}{4.5pt}
\renewcommand{\arraystretch}{1}
\label{tab:perturbation}
\begin{tabular}{@{}l||c||c||c||c@{}}
\hline
Judge & $A_j$ & \shortstack{$A_j$, semantic\\arm in loops} & \shortstack{Reactions,\\sem.\ / ctl.} & $\rho_j$ \\
\hline \hline
OpenAI & \shortstack{0.8655\\{\scriptsize[0.824, 0.905]}} & \shortstack{0.9465\\{\scriptsize[0.905, 0.980]}} & \shortstack{113/144\\11/92} & \shortstack{0.2717\\{\scriptsize[0.124, 0.405]}} \\ \hline
DeepSeek & \shortstack{0.8478\\{\scriptsize[0.814, 0.881]}} & \shortstack{0.9041\\{\scriptsize[0.867, 0.941]}} & \shortstack{225/296\\27/180} & \shortstack{0.3159\\{\scriptsize[0.216, 0.412]}} \\ \hline
Gemini & \shortstack{0.7512\\{\scriptsize[0.719, 0.783]}} & \shortstack{0.8438\\{\scriptsize[0.796, 0.891]}} & \shortstack{161/287\\14/179} & \shortstack{0.3681\\{\scriptsize[0.271, 0.452]}} \\ \hline
Claude & \shortstack{0.8380\\{\scriptsize[0.806, 0.871]}} & \shortstack{0.8982\\{\scriptsize[0.853, 0.939]}} & \shortstack{137/200\\1/129} & \shortstack{0.2464\\{\scriptsize[0.132, 0.359]}} \\
\hline \hline
\end{tabular}
\end{table}

\subsubsection{Arm Discrimination and Specificity}
All four judges reliably separate a broken program from an edited one that behaves the same, every interval lying above 0.5 (~\tabboxref{tab:perturbation}). OpenAI, DeepSeek and Claude form one tier, between 0.8380 and 0.8655 with overlapping intervals, and Gemini sits below them, its interval clear of all three, at 0.7512 [0.719, 0.783]. The arms differ in size as well as in rank: semantic mutants lower the judged probability by 0.49 to 0.69 on average, control mutants by 0.004 to 0.092. Specificity distinguishes the judges more sharply still. Claude is the most specific, reacting to 1 of its 129 control mutants against 27 of 180 for DeepSeek, and on the post-increment rewrite, which provably changes nothing, DeepSeek's score falls by 0.124 on average where Claude's rises by 0.001.

\subsubsection{Grading by Severity}
\begin{figure}[h]
\centering
\includegraphics[width=3in]{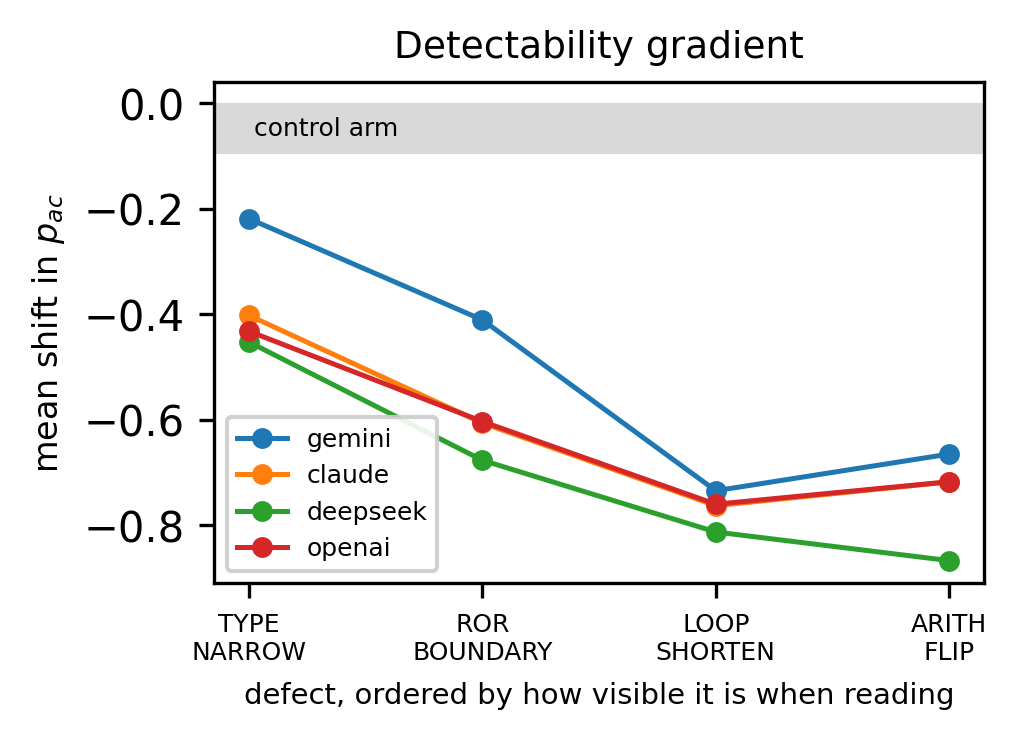}
\caption{Mean shift in the judged probability per semantic operator, ordered by severity, primary subset, \textsc{blind} condition. The grey band spans the control arm's mean shifts}
\label{fig_pert_gradient}
\end{figure}

Every judge's drop grows with the severity of the defect, with $\rho_j$ from 0.2464 [0.132, 0.359] for Claude to 0.3681 [0.271, 0.452] for Gemini, every interval excluding zero (\figboxref{fig_pert_gradient}). The subtlest defect separates the judges most: on a narrowed integer type, which still compiles and stays correct below 32 bits, Gemini's score falls by 0.22 on average against 0.40 to 0.45 for the others. The gradient is steepest across the subtler defects and levels off at the top, where the most visible defect, the arithmetic flip, already draws drops of 0.67 to 0.87.

\subsubsection{Two Abilities, Not One}
\begin{figure}[h]
\centering
\includegraphics[width=3in]{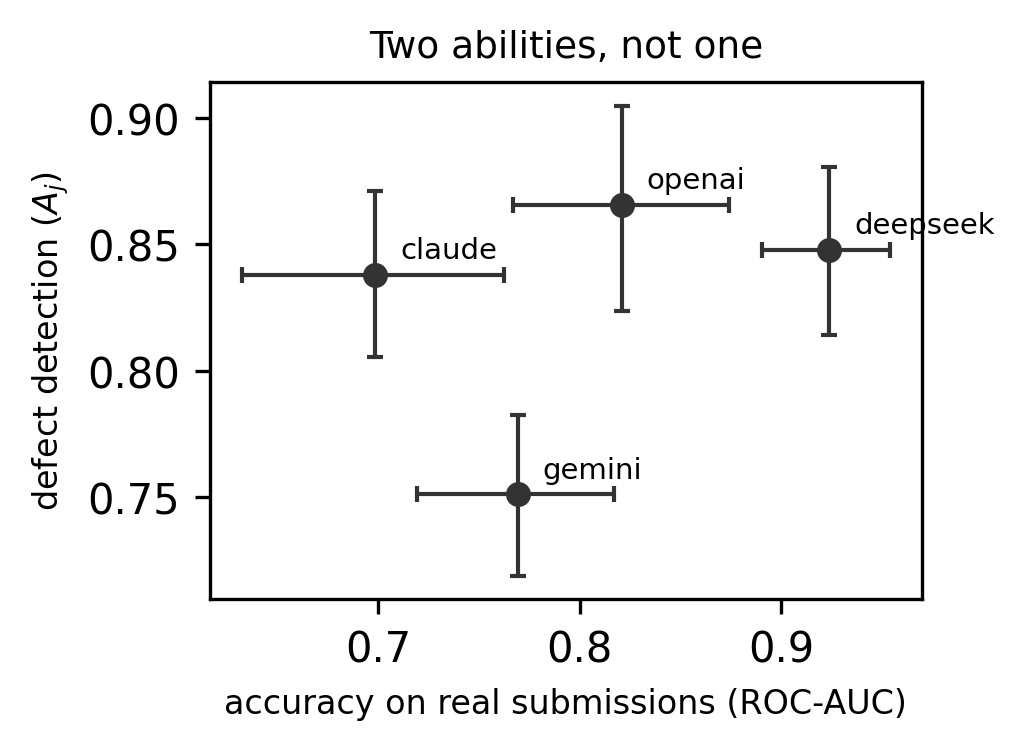}
\caption{Accuracy (ROC-AUC, \textsc{nonblind}, 517 attempts) against defect detection ($A_j$, \textsc{blind}, primary subset), with 95\% intervals}
\label{fig_pert_two_abilities}
\end{figure}

Ranking real submissions and reacting to a broken one are separate abilities (\figboxref{fig_pert_two_abilities}). Claude and Gemini cannot be told apart on accuracy, 0.6985 [0.633, 0.763] against 0.7691 [0.719, 0.817], yet separate on detection, 0.8380 [0.806, 0.871] against 0.7512 [0.719, 0.783]. DeepSeek, first on accuracy (0.9236 [0.890, 0.954]), is not first on detection, where it overlaps OpenAI and Claude. With four judges, no correlation between the two abilities is computed; the separation rests on the intervals alone.

\subsubsection{Conservative Estimates}
A mutant's defect is fixed by construction, since the contest's hidden tests run only inside its official judge system, so any intended defect that happens not to change behavior counts against a judge as a missed detection; and 35 of the 348 semantic mutants sit in a guard context, where the edit may be unreachable under the problem's own constraints. Both can only lower $A_j$, never raise it. Setting the guard mutants aside, by restricting the semantic arm to loop contexts, raises every judge's $A_j$ to between 0.8438 for Gemini and 0.9465 for OpenAI (~\tabboxref{tab:perturbation}), in line with some guard mutants being behaviorally equivalent; since the restriction also sets aside the subtlest operator, which never occurs in a loop, the two effects are not separated here. The result also holds on the full set: on every scored mutant, including seeds a judge misjudged, every interval excludes 0.5, from 0.7359 [0.706, 0.768] for Gemini to 0.8246 [0.792, 0.857] for DeepSeek.

\subsection{Human-AI Co-Creation Dynamics Results}
\label{co-creation}
The co-creation analyses fall into two groups, reported in that order. Success@Turn, time-to-success, revision churn, and code-aware similarity are computed from submissions and verdicts alone and do not depend on any judge. The turn-wise rank consensus $\bar q_{u,p,k}$ and the judged improvement $\Delta \bar q_{u,p,k}$ (Section~\ref{sec:approach_analysis}) do depend on the judges; they are computed under \textsc{nonblind} and rest on the validated discrimination of the four-judge panel whose score they use, ROC-AUC 0.8734 [0.831, 0.913] (Section~\ref{results_judge}), so that no co-creation finding rests on one model.

\subsubsection{Success by Turn and Time-to-Success}
\begin{figure*}[t]
\centering
\includegraphics[width=0.9\textwidth]{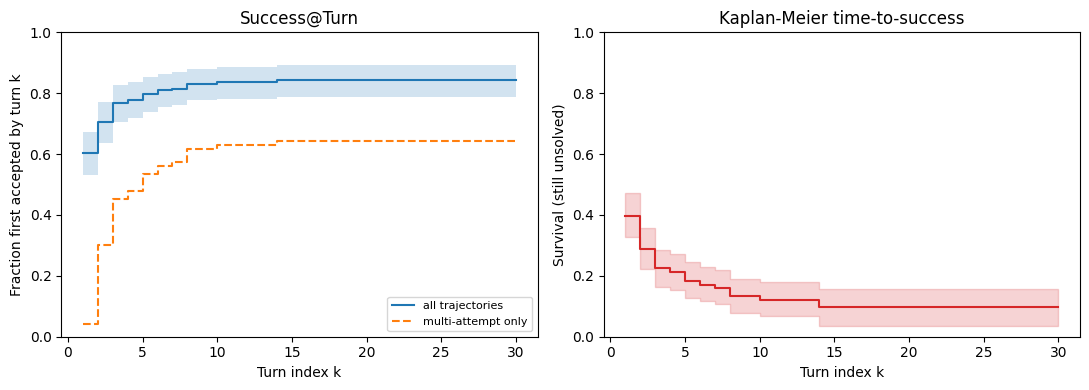}
\caption{(a) Success@Turn, the share of trajectories first accepted by turn $k$: all 184 (solid, 95\% interval) and the 73 multi-attempt ones (dashed). (b) Kaplan-Meier share still unsolved, never-solved trajectories censored at their last attempt}
\label{success_by_turn}
\label{kaplan_meier}
\end{figure*}

Success accumulates across turns (\figboxref{success_by_turn}): Success@Turn rises from 0.6033 [0.533, 0.674] at the first attempt to 0.7065 [0.636, 0.772] by the second and 0.8370 [0.783, 0.886] by the tenth, toward the 0.8424 of trajectories ever solved (155 of 184). The first-turn value carries a selection effect that has to be read with it: a trajectory is single-attempt largely because its first attempt succeeded, so 108 of the 111 single-attempt trajectories are solved against 47 of the 73 multi-attempt ones. The conditional quantity is the informative one. Of the 73 trajectories not solved at their first attempt, 44 (60.3\%) succeed later, 44 of the 70 that continued (62.9\%), with the first success at a median of turn 3 and as late as turn 14. For the multi-attempt trajectories alone, Success@Turn climbs from 0.041 to 0.644, so most of their success comes after the first turn. The Kaplan-Meier curve, which right-censors the 29 never-solved trajectories at their last attempt, gives the same picture from the other side: survival stands at 0.397 after the first turn, declines through the early turns, and is flat after turn 14.

\subsubsection{Revision Magnitude and Similarity to the Accepted Solution}
\begin{figure}[h]
\centering
\includegraphics[width=2.8in]{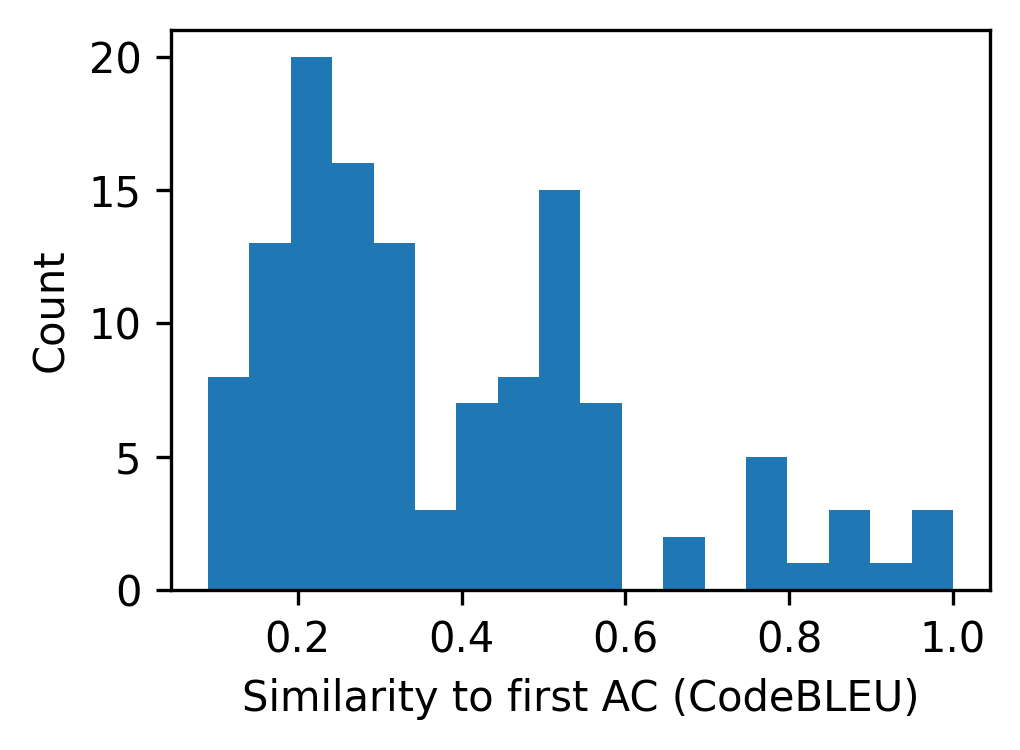}
\caption{CodeBLEU similarity to the first accepted submission in solved trajectories: 125 of 280 observations, after excluding the 155 that score that submission against itself}
\label{CodeBLEU_convergence}
\end{figure}

Revisions between consecutive attempts are substantial: across the 333 transitions of the 73 multi-attempt trajectories, the NED churn has a median of 0.5138. Revision size does not distinguish the transitions that reached acceptance from those that did not, their mean churn differing by +0.0406 [$-0.032$, +0.111]. Among solved trajectories, the CodeBLEU similarity of each attempt to the first accepted submission, with that submission excluded from its own comparison, has a median of 0.3103 and a first quartile of 0.2109 over 125 observations, with 6.4\% above 0.8 (\figboxref{CodeBLEU_convergence}); the one value at 1.0 is an attempt that resubmitted the accepted code unchanged. The distribution stands as a descriptive reference: comparing it with the similarity of independent solutions to the same problems is the step that would turn it into a test of convergence.

\subsubsection{Turn-Wise Rank Consensus}
\begin{figure*}[t]
\centering
\includegraphics[width=0.9\textwidth]{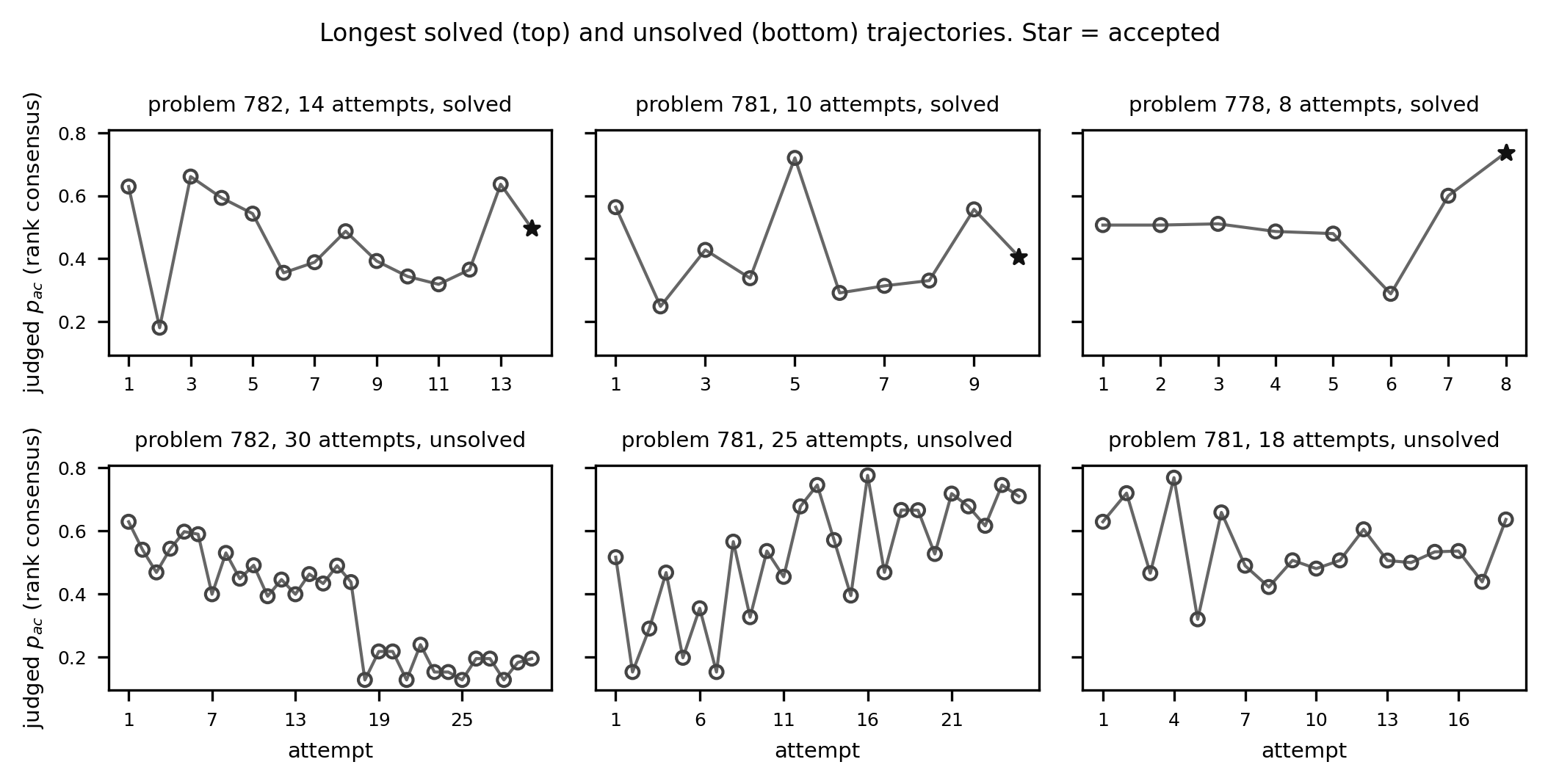}
\caption{Rank consensus $\bar q$ per attempt, \textsc{nonblind}, for the three longest solved (top) and three longest unsolved (bottom) trajectories; a star marks acceptance. Long trajectories cluster on the hardest problems, so the panel is not a random sample}
\label{struggle_curve}
\end{figure*}

\begin{figure}[h]
\centering
\includegraphics[width=2.9in]{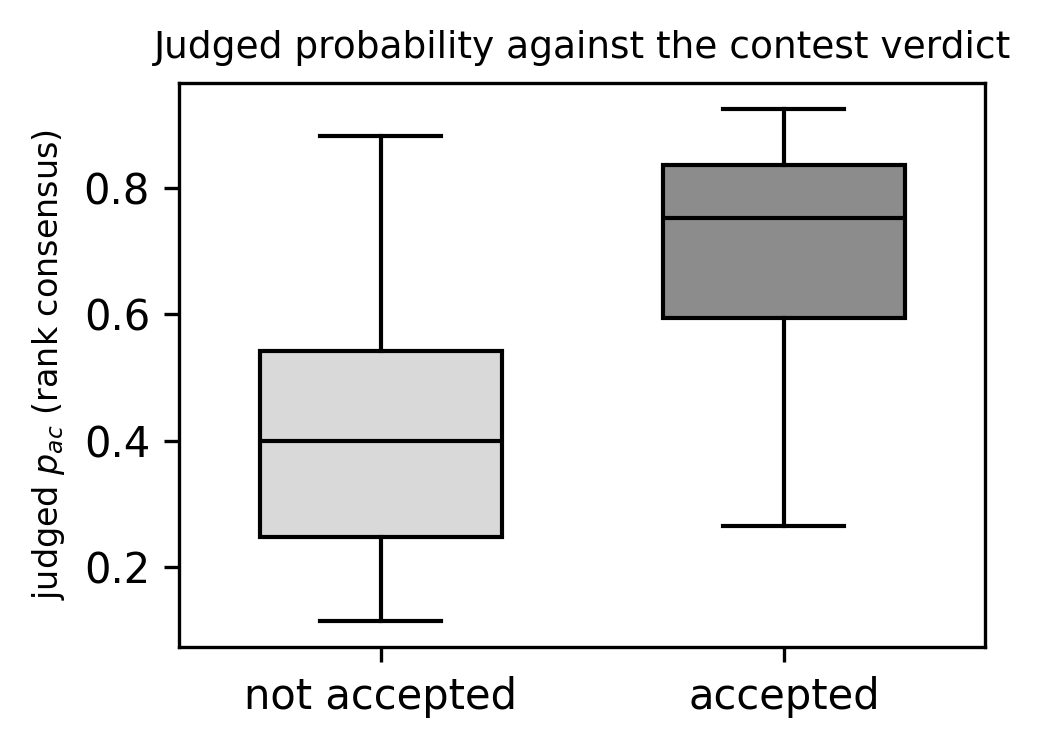}
\caption{Rank consensus $\bar q$ by verdict, \textsc{nonblind}: 359 non-accepted and 158 accepted attempts}
\label{fig_pac_by_outcome}
\end{figure}

The six trajectories in~\figboxref{struggle_curve} follow the fixed selection rule in its caption. The solved ones reach acceptance after uneven paths of 8 to 14 attempts. The unsolved ones fall on problems 781 and 782, and problem 781 drew a single acceptance in 115 attempts across all participants, the one in the top row; one of them rises past a rank consensus of 0.7 without ever being accepted, so a rising curve does not by itself imply acceptance. Across all 517 attempts, the rank consensus separates the verdicts clearly (\figboxref{fig_pac_by_outcome}): its median is 0.40 for the 359 non-accepted attempts and 0.75 for the 158 accepted, with interquartile ranges of [0.25, 0.54] and [0.59, 0.84] that do not overlap, while the whiskers do.

\subsubsection{Revision Magnitude Against Judged Improvement}
\begin{figure}[h]
\centering
\subfloat[NED churn]{\includegraphics[width=1.74in]{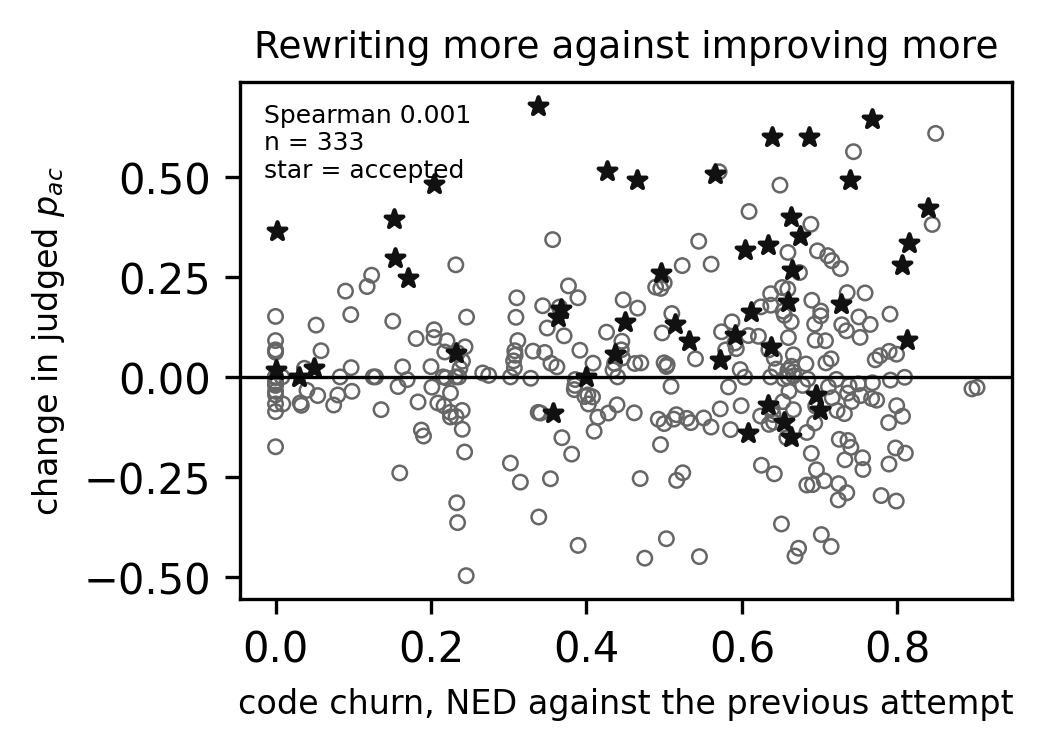}\label{ned_based_code_churn}}
\hfil
\subfloat[CodeBLEU distance]{\includegraphics[width=1.74in]{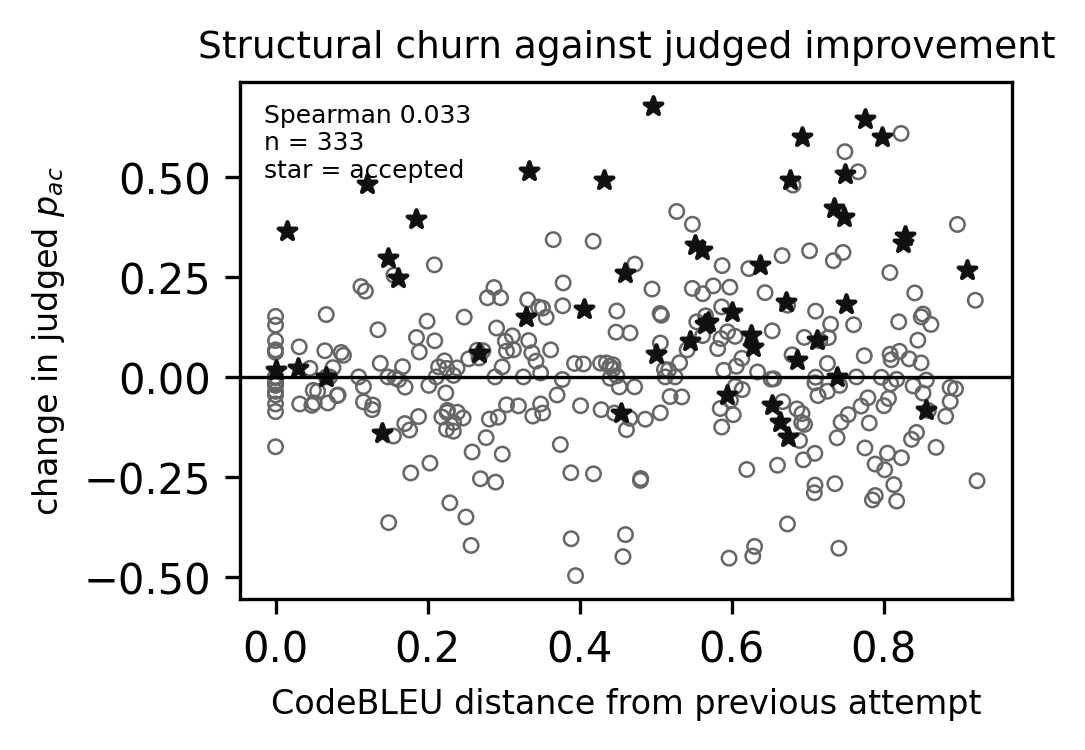}\label{CodeBLEU_code_churn}}
\caption{Revision size between consecutive attempts against judged improvement $\Delta \bar q$, 333 transitions, \textsc{nonblind}; stars mark transitions reaching acceptance}
\label{fig_churn_pair}
\end{figure}

Larger revisions do not bring larger judged gains (\figboxref{fig_churn_pair}). Over the same 333 transitions, NED churn and the judged improvement $\Delta \bar q$ are uncorrelated, Spearman 0.0007 [$-0.085$, +0.090], with the interval clustered on the 73 trajectories, and the code-aware CodeBLEU distance agrees at 0.033, for which no clustered interval was computed. With the judge-independent result above, the finding holds on both sides of the measurement: revision size predicts neither the judged improvement nor the verdict. Accepted transitions appear across the whole range, from small targeted edits to near rewrites, so productive revision in this setting takes more than one form.

\subsection{Synthesis with respect to RQs and EQs}
\label{sec:rq_eq_synthesis}
Each pair is answered evaluation question first, since a research question's answer is admissible only where its evaluation question holds; EQ3 comes first because the judge-dependent parts of RQ1 and RQ2 rest on it.

\subsubsection{EQ3/RQ3: Measurement quality and judges as measurement instruments}
EQ3 is met on all five criteria, with limits stated per judge. Every judge discriminates against the contest verdict, every interval excluding chance, also under problem clustering (\tabboxref{judge_all_metrics}). Out-of-fold thresholds transfer intact for Gemini and nearly so for DeepSeek, while OpenAI and Claude lose 62.5\% and 45.7\% of their in-fold MCC under \textsc{blind}, on scales with as few as 8 distinct values (Section~\ref{sec:scale_validity}). Calibration separates DeepSeek and OpenAI from Claude and Gemini; agreement is beyond chance for every pair and higher with relevance-selected context than with none (\tabboxref{kappa_summary}); and every judge's score falls further on a broken program than on a behavior-preserving edit, every severity gradient excluding zero (\tabboxref{tab:perturbation}). RQ3 is answered on that basis. Multiple LLM judges provide auditable assessments whose stability is measured: agreement is moderate, Fleiss' $\kappa$ 0.4353 [0.366, 0.510] under \textsc{nonblind}, and depends on the request, since supplying the problem statement rebuilt the judges' ordering (rank correlation 0.3021 [0.168, 0.424]). The judges are therefore not interchangeable. DeepSeek ranks attempts best, Claude reacts least to edits that change nothing, and Claude and Gemini, indistinguishable on accuracy, separate on defect detection, so the choice of judge follows its intended use.

\subsubsection{EQ1/RQ1: Trajectory validity and co-creation dynamics}
EQ1 is met: Success@Turn, with intervals, and time-to-success under right-censoring characterize trajectories beyond their final attempt from verdicts alone (\figboxref{success_by_turn}). RQ1's answer is that success accumulates across turns: 44 of the 73 trajectories unsolved at the first attempt (60.3\%) succeed later, at a median of turn 3, a result an outcome-only evaluation cannot reach. The judges' rank consensus, which rests on EQ3 through the panel's ROC-AUC of 0.8734 [0.831, 0.913], separates accepted from non-accepted attempts (medians 0.75 and 0.40, \figboxref{fig_pac_by_outcome}) and traces uneven paths in which a rise does not by itself imply acceptance.

\subsubsection{EQ2/RQ2: Churn and similarity validity, and revision behavior}
EQ2 is met: NED churn and CodeBLEU distance are computed for all 333 transitions, and the relation of NED churn to progress carries an interval. The earlier convergence reading is withdrawn: similarity to the first accepted solution, median 0.3103 over 125 observations (\figboxref{CodeBLEU_convergence}), stays a description until a baseline of independent solutions turns it into a test. RQ2's answer rests on one finding: revision size does not predict progress. It is unrelated to judged improvement, Spearman 0.0007 [$-0.085$, +0.090], and does not separate the transitions that reached acceptance, +0.0406 [$-0.032$, +0.111] (\figboxref{fig_churn_pair}), so incremental refinement and broader restructuring both lead to acceptance. The verdict side involves no judge and reaches the same null, so the finding does not rest on the judges alone; on the judged side, which rests on EQ3, the null is also compatible with reports that LLM judges are insensitive to refinement effort~\cite{wang2025axiom}.

\section{Discussion}
\label{Dis}
\subsection{Overall interpretation of the findings}
Co-creation progress continues well beyond the first attempt: success keeps accumulating across turns, and most trajectories that miss at first go on to succeed. Early success rates also carry a selection effect, since a trajectory is short largely because its first attempt succeeded; the conditional view, success after an initial failure, removes that confound and shows persistence paying off. Revision size predicts neither judged improvement nor the verdict, so progress shows in what a revision achieves, not in how much it changes. On the measurement side, ranking quality, calibration, decision stability, and defect sensitivity order the judges differently, so each criterion guides a different choice of judge. Execution outcomes remain the anchor; attempt-level tracking with a validated instrument extends what they reveal about co-creation.

\subsection{LLM-as-a-Judge as an evaluation framework}
Rubric-driven LLM judges serve as measurement instruments for co-creative coding when their outputs are treated as structured, auditable signals, and turning the pipeline on its own outputs yields three findings that make this practical. First, the specification of the judge input is a first-order design variable: supplying the problem statement rebuilt the judges' ordering, so judge performance is reported together with the request that produced it, given here for each condition with its coverage evidence. Second, scale validity is measured alongside schema compliance: it shows how many distinct decisions each judge's probabilities support, and range-bounded output checks now catch out-of-range rubric scores of the kind an unbounded schema check let through on 39 rows of an earlier run. Third, accuracy and defect sensitivity are measured as distinct properties, so each judge is matched to the use it has been validated for. Because the judges are not interchangeable, the framework states which judge suits which use, on the evidence of Sections~\ref{results_judge} and~\ref{sec:perturbation_results}. The pipeline is auditable end to end. Context provenance is stated per request; every row is judged on the full input, and any row that cannot be is recorded as a failure, never shortened to fit; and model identity is recorded as the provider reports it, which matters because one requested identifier resolved to a different served model.

\subsection{Human-AI co-creation in coding and software engineering}
The findings divide into those that transfer and those specific to the contest, and both are useful. The three judge-side findings are likely to transfer to other uses of an LLM as a judge, since they concern how a judge's output relates to its input, its scale and its sensitivity. The apparatus transfers as well: attempt-level trajectories, conditional success, revision-magnitude measures and the perturbation check apply wherever attempts are logged against some ground truth. The co-creation outcomes transfer as patterns, not as numbers, since they reflect competitive programming under time pressure. One pattern agrees with practice: progress follows no single revision style, so effective co-creation rests on the participant's capacity to evaluate intermediate outputs and steer the next revision, a capacity that grows more valuable as AI assistance improves.

\subsection{Practical implications for future co-creation studies}
Several practices follow. Log every attempt, since conditional success and time-to-success depend on it; report the judge request, including what the judge sees and where its context is cut; check a judge's output scale before relying on its threshold, and record the served model identity with every response; choose each judge for its intended use, with a panel ranked on a common scale as a safe default when no single judge is validated for that use; and measure defect sensitivity directly, with a perturbation check against matched controls.

\subsection{Limitations}
The findings hold within six boundaries, each opening a natural extension.
\textbf{(i) Scope:} one contest, with 15 participants, 13 problems and 517 attempts; the judge-side findings are likely to transfer, as they concern how a judge's output relates to its input, scale and sensitivity, while the co-creation patterns invite replication in open-ended development.
\textbf{(ii) Context:} participant context is reconstructed from logs, so context-dependent results are also reported on the evidence-backed subset.
\textbf{(iii) Hosted judges:} identifiers can be aliases and outputs can vary at temperature zero, so the served identity is recorded with every response.
\textbf{(iv) Perturbation:} defects are intended, not executed, and some guard-context edits may be unreachable, so the reported sensitivity is a conservative estimate.
\textbf{(v) Revision and similarity:} a baseline of independent solutions would turn the similarity distribution into a convergence test, and the judged-side null may partly reflect judge insensitivity, although the verdict side shows the same null.
\textbf{(vi) CodeBLEU:} the data-flow component is excluded and C is scored under the C++ configuration, which affects only the post-hoc similarity analysis.

\section{Conclusion}
\label{Cn}
This study presented a reliability-aware \emph{LLM-as-a-Judge} framework for evaluating \emph{human-AI co-creation} in contest-style coding, in which four judges score every attempt through a verified, schema-constrained pipeline under three input conditions, and their measurement quality is tested alongside the co-creation analysis.
 
Four findings stand out. First, the judge input is a first-order design variable: supplying the problem statement rebuilt the judges' ordering of the same attempts (rank correlation 0.3021 [0.168, 0.424]) and raised out-of-fold ROC-AUC by 0.130 [0.061, 0.202] to 0.295 [0.214, 0.381] for the three judges whose model was unchanged. Second, schema compliance does not guarantee a usable scale: the judges emit as few as 8 distinct probability values, leaving as few as 6 effective decision rules on a 99-point threshold grid. Third, accuracy and defect sensitivity are separable: every judge detects injected defects, yet Claude and Gemini, indistinguishable on accuracy (0.6985 [0.633, 0.763] and 0.7691 [0.719, 0.817]), separate on detection (0.8380 [0.806, 0.871] and 0.7512 [0.719, 0.783]), so each judge is chosen for its use. Fourth, success accumulates across turns, from 0.6033 [0.533, 0.674] at the first attempt to 0.8370 [0.783, 0.886] by the tenth; 44 of the 73 trajectories unsolved at first (60.3\%) succeed later, and revision size predicts neither judged improvement (Spearman 0.0007 [$-0.085$, +0.090]) nor the verdict.
 
Future work follows directly: weighted or selected judge panels, set against the unweighted panel examined here; a baseline of independent solutions, which would turn the similarity analysis into a test of convergence; perturbation checks with participant context and execution-confirmed defects; and replication beyond contest-style programming. Together, the results show that execution outcomes, tracked across attempts and read through validated judges, reveal both how co-creative solutions form and how far a judge's assessment can be trusted.

\bibliographystyle{ieeetr}

\bibliography{Main}

\vfill

\end{document}

%% file: Graphs_Figures/comp1.tex
\begin{tikzpicture}

\begin{axis}[
  ybar,
  bar width=3.2pt,
  width=\linewidth,
  height=4.8cm,
  ymin=0, ymax=1.00, 
  ylabel={Discrimination (AUC)},
  ylabel style={yshift=-6pt}, 
  symbolic x coords={OpenAI,DeepSeek,Gemini,Claude},
  xtick=data,
  xticklabel style={rotate=25, anchor=east},
  tick label style={font=\scriptsize},
  label style={font=\small},
  legend style={
    font=\scriptsize,
    at={(0.5, -0.25)},
    anchor=north,
    legend columns=3
    column sep=4pt,
    draw=black,
    fill=none
  },
  ymajorgrids=true,
  grid style={opacity=0.25},
  enlarge x limits=0.18
]
\addplot [color=orange, fill=orange!20, semithick, pattern=crosshatch, pattern color = orange] coordinates {
  (OpenAI,0.8208)
  (DeepSeek,0.9236)
  (Gemini,0.7691)
  (Claude,0.6985)
};

\addplot [color=blue, fill=blue!20, semithick, pattern=grid, pattern color = blue] coordinates {
  (OpenAI,0.6969)
  (DeepSeek,0.8844)
  (Gemini,0.5480)
  (Claude,0.5060)
};

\addlegendimage{black,thick,mark=diamond*,mark options={fill=red}}

\legend{ROC-AUC, PR-AUC, $\mathrm{MCC}_{\mathrm{oof}}$} 
\end{axis}

\begin{axis}[
  width=\linewidth,
  height=4.5cm,
  ymin=0, ymax=1.00, 
  axis y line*=right,
  axis x line=none,
  ylabel={$\mathrm{MCC}_{\mathrm{oof}}$}, 
  symbolic x coords={OpenAI,DeepSeek,Gemini,Claude},
  xtick=\empty,
  tick label style={font=\scriptsize},
  label style={font=\small},
  ylabel style={yshift=6pt}, 
]
\addplot[
  thick,
  color=black,
  mark=diamond*,
  mark options={fill=red},
  sharp plot,
  mark size=2.2pt
] coordinates {
  (OpenAI,0.4396)
  (DeepSeek,0.7822)
  (Gemini,0.4978)
  (Claude,0.2120)
};

\end{axis}

\end{tikzpicture}

%% file: Graphs_Figures/comp2.tex
\begin{tikzpicture}
\begin{axis}[
  ybar,
  bar width=3.2pt,
  width=\linewidth,
  height=4.8cm,
  ymin=0, ymax=1.05,
  ylabel={Normalized quality},
  ylabel style={yshift=-6pt},    
  symbolic x coords={OpenAI,DeepSeek,Gemini,Claude},
  xtick=data,
  xticklabel style={rotate=25, anchor=east},
  tick label style={font=\scriptsize},
  label style={font=\small},
  legend style={
    font=\scriptsize,
    at={(0.5, -0.26)},
    anchor=north,
    legend columns=3
    column sep=4pt,
    draw=black,
    fill=none
  },
  ymajorgrids=true,
  grid style={opacity=0.25},
  enlarge x limits=0.20
]
\addplot [color=teal, fill=teal!20, semithick, pattern=crosshatch dots, pattern color = teal] coordinates {
  (OpenAI,0.7368)
  (DeepSeek,1.0000)
  (Gemini,0.0000)
  (Claude,0.1615)
};

\addplot [color=black, fill=black!20, semithick, pattern=horizontal lines, pattern color = black] coordinates {
  (OpenAI,0.9893)
  (DeepSeek,1.0000)
  (Gemini,0.0000)
  (Claude,0.4979)
};

\addplot [color=green!50!black, fill=green!20, semithick, pattern=north east lines, pattern color = green!50!black] coordinates {
  (OpenAI,0.5898)
  (DeepSeek,1.0000)
  (Gemini,0.0000)
  (Claude,0.1022)
};

\legend{ECE$_{15}$, $\mathcal{L}$, $B$}
\end{axis}
\end{tikzpicture}

%% file: Graphs_Figures/comp3.tex
\begin{tikzpicture}
\begin{axis}[
    width=0.95\linewidth,
    scale only axis,
    height=4.8cm,
    ybar,
    bar width=5pt,
    ylabel={Raw score},
    symbolic x coords={Claude,DeepSeek,Gemini,OpenAI},
    xtick=data,
    x tick label style={rotate=25,anchor=east},
    ymin=0, ymax=0.85, 
    grid=major,
    enlarge x limits=0.18,
    tick label style={font=\normalsize},
    label style={font=\normalsize},
    legend style={
        at={(0.5, -0.40)},
        anchor=south,
        legend columns=1, 
        draw=black,
        fill=none,
        font=\normalsize
    },
    nodes near coords={\pgfmathprintnumber[fixed,precision=2]{\pgfplotspointmeta}},
    point meta=y,
    every node near coord/.append style={
        font=\scriptsize,
        rotate=90,
        anchor=west
    }
]


\addplot [color=blue!60, fill=black, semithick, pattern=crosshatch dots, pattern color = blue!60] coordinates {
    (Claude,   0.2120)
    (DeepSeek, 0.7822)
    (Gemini,   0.4978)
    (OpenAI,   0.4396)
};


\addplot [color=darkgray, fill=black, semithick, pattern=bricks, pattern color = darkgray] coordinates {
    (Claude,   0.4102)
    (DeepSeek, 0.3932)
    (Gemini,   0.4661)
    (OpenAI,   0.4769)
};

\legend{$\mathrm{MCC}_{\mathrm{oof}}$, Mean pairwise $\kappa$} 

\end{axis}
\end{tikzpicture}

%% file: Graphs_Figures/comp4.tex
\begin{tikzpicture}
\begin{axis}[
    width=0.95\linewidth,
    scale only axis,
    height=5cm,
    ybar,
    bar width=5pt,
    ylabel={Normalized quality score},
    symbolic x coords={Claude,DeepSeek,Gemini,OpenAI},
    xtick=data,
    x tick label style={rotate=25,anchor=east},
    ymin=0, ymax=1.12, 
    grid=major,
    enlarge x limits=0.12,
    tick label style={font=\normalsize},
    label style={font=\normalsize},
    legend style={
        at={(0.5, -0.40)},
        anchor=south,
        legend columns=1, 
        draw=black,
        fill=none,
        font=\normalsize
    },
    nodes near coords={\pgfmathprintnumber[fixed,precision=2]{\pgfplotspointmeta}},
    point meta=y,
    every node near coord/.append style={
        font=\scriptsize,
        rotate=90,
        anchor=west
    }
]

\addplot [color=black, fill=black, semithick, pattern=north east lines, pattern color = black] coordinates {
    (Claude,   0.0000)
    (DeepSeek, 1.0000)
    (Gemini,   0.5012)
    (OpenAI,   0.3992)
};

\addplot [color=purple, fill=black, semithick, pattern=crosshatch, pattern color = purple] coordinates {
    (Claude,   0.2031)
    (DeepSeek, 0.0000)
    (Gemini,   0.8710)
    (OpenAI,   1.0000)
};

\legend{Relative $\mathrm{MCC}_{\mathrm{oof}}$, Relative mean pairwise $\kappa$} 

\end{axis}
\end{tikzpicture}